\documentclass[12pt,preprint]{aastex}

\newcommand{\rb}{r_{\rm break}}
\newcommand{\re}{r_{\rm distort}}
\newcommand{\rt}{r_{\rm tide}}
\newcommand{\rtp}{r_{\rm tide, peri}}
\newcommand{\rti}{r_{\rm tide, inst}}

\begin{document}

\title{Interpreting the Morphology of Diffuse Light Around
Satellite Galaxies}

\author{Kathryn V. Johnston\\ Van Vleck Observatory, Wesleyan
University, Middletown, CT 06459\\ E-mail: kvj@astro.wesleyan.edu}

\author{Philip I. Choi, Puragra Guhathakurta\\ UCO/Lick Observatory,
University of California, Santa Cruz, CA 95064\\ E-mail:
pchoi@ucolick.org, raja@ucolick.org}

\begin{abstract} 

Recent observations of surface brightness distributions of both Milky
Way and M31 satellite galaxies have revealed many instances of sudden
changes or breaks in the slope of the surface brightness profiles (at
some break radius, $\rb$).  These breaks are often accompanied by
increasingly elliptical isophotes and sometimes by isophote twisting.
We investigate the hypothesis of a tidal origin for these features by
applying the same ellipse fitting techniques that are used on observed
galaxies to numerical simulations of the destruction of
satellites, represented by spherical, single-component systems.
We examine how observed quantities such as $\rb$, ellipticity $e$ and
position angle $\phi$ of the fitted ellipses and amplitude of
extra-break population vary with the satellite's orbital eccentricity
and phase, and our viewpoint relative to the orbit.  We also look at
orbit and viewpoint dependence of the rate of change of the latter
three quantities with radius.

We find that there are trends with orbital phase and
eccentricity in all observed  quantities, many of which are preserved 
through a wide variety of viewing angles.
In particular, a generic feature of all simulations is a depletion zone
just interior to an excess zone, regions in which the surface brightness
is lower and higher, respectively, than the initial profile.
A clear interpretation of any individual image, however, is likely to be 
hampered by the dependence of the observable features on these multiple
parameters.
For example, breaks can be excited by several physical processes
and can occur well within the bound satellite population.
Nevertheless, we do find we can
place loose constraints on the 
tidal radius, mass loss rate, 
orbital type and phase of the satellite, and nature of breaks 
using photometric data alone. 
\end{abstract}

\clearpage

\section{Introduction}

The surface brightness distributions of Galactic globular
clusters and dwarf spheroidal satellites have traditionally been studied
using star count data. 
In the last few years, very low 
surface brightness levels have been reached by removing contaminants
either by making use of a survey that covers a significant
area around the satellite so the background counts
can be assessed \citep{ih95},
making some color cuts \citep{ggrf,g+95,lmc00} or concentrating on specific
classes of associated stars such as RR Lyrae \citep{ksh96} or
metal-poor giants \citep{m+00}.
These approaches have revealed many instances of the slope of a satellite's
surface brightness profile abruptly changing at some radius $\rb$,
so that the satellite is 
apparently surrounded by a diffuse envelope of stars
(see Grillmair, 1998, for a more complete listing).
In some instances, this break is accompanied by 
distortions and twisting of the star count isopleths, and even
hints of a stream of debris extending along a particular
axis \citep{lmc00,sdss}.

\citet{g+96} report indications of tidal debris around some of M31's
globular clusters using HST to perform a similar star count analysis.
However, to reach such low surface brightness levels around satellites
of external galaxies in general will mean working in integrated light
rather than individual star counts.  We can get a sense for the
sensitivity required by noting that debris found using star counts
separated from Sagittarius by 10--40 degrees has been estimated to
correspond to steadily decreasing surface brightnesses in the range
$\mu_V\sim27\>$--30\,mag~arcsec$^{-2}$ \citep{mom98}.  Clearly, the
surface brightness of debris will depend on the mass loss rate and
orbit of the satellite --- for example, the giant star count densities
reported recently in \citet{m+00} around Carina correspond to a drop
of roughly a factor of 100 from their central value, or $\mu_V\sim
30$\,mag arcsec$^{-2}$.  In \citet{paper2} we present the results of
work where we find low surface brightness features similar to the
extra-break populations of the Milky Way's satellites around
satellites of M31: NGC\,205 and M32.  This was made possible by a
wide-field CCD image of M31, a mosaic of 60 precisely flat-fielded
and sky-subtracted frames \citep{gcr01}, which allowed the careful
subtraction of M31's light from the fields around NGC\,205 and M32.  This
approach allowed for the extension of work by \citet{hodge73} and
\citet{kent87} to surface brightness limits of $\mu_B\sim 27$\,mag
arcsec$^{-2}$.

These ``extra-break'' features are reminiscent of 
particle distributions seen in simulations of the destruction of a
satellite in a tidal field \citep{g+95,jsh99,clm99}.
Clearly this work is closely related to the vast literature 
on signatures of mergers of galaxies such as tidal tails 
\citep{tt72} and shells \citep{q84}.
The advantage of looking at satellite galaxies is that their
influence on their parent is negligible and this simplifies
the dynamics of the debris considerably.
Once lost from the satellite, debris in these simulations is seen
to spread along the orbit of the satellite in thin streams
at a predictable rate \citep{t93,j98,hw99}.
This has led to the suggestion that the morphology of
extra-break populations can
be used to constrain the direction in which the satellite is moving
\citep{g+95,clm99}, which in turn could tell us something about
the satellite's orbit --- information that is useful
for both understanding the satellite's history and for comparing orbital
distributions of a galaxy's satellite system with cosmological
models \citep{tormen97,ghigna98}.
Other work has shown that the surface density of
the debris can be used to estimate the mass-loss rate from 
the satellite \citep{jsh99}, and that the limitation of the satellite
or extent of debris gives some indication of the satellite's mass and
hence dark matter content
\citep{fl83,km89,k93,m96,burkert97,kk98}. 

In light of this steady stream of results on extra-break populations
from within and beyond the Milky Way, the aim of this paper is to
confirm and generalize some of the theoretical ideas of earlier
studies. In particular, we are interested not in how or why a satellite is
disrupting, but rather in what general statements we can make
about its state and past evolution 
by observing its debris.  Our current intuition is that: (i)
the physical scale (corresponding to $\rb$) over which debris is
distributed is proportional to the ratio $(m_{\rm sat}/M_{\rm
gal})^{1/3}$, where $m_{\rm sat}$ is the satellite's mass and
$M_{\rm gal}$ is the parent galaxy mass enclosed within the 
pericenter of the satellite's orbit \citep{j98}; (ii) as long as we scale the
size of any box we are considering around the satellite by this ratio,
the morphology (i.e. two-dimensional shape and one-dimensional
profile) of the extra-break population will depend only on orbital
eccentricity, orbital phase and viewpoint; and (iii) the amplitude of
the surface-density of the extra-break population is proportional to
the mass-loss rate from the satellite \citep{jsh99}.  The mass loss rate
itself will vary with the orbit and density structure of the satellite
\citep{w94a,w94b,w94c,jhb96},
but we do not study these dependencies directly since they
will only affect the amplitude and not the overall morphology of the
extra-break population. 
We instead focus on hypotheses (ii) --- 
the dependencies that have been explored the least ---
by looking at simulations of satellites being
destroyed along orbits with a variety of eccentricities and viewed at
all orbital phases along many directions.

We have a 
particular interest in the appearance of
companions to external galaxies, at sufficiently
large distance from their parents that their surroundings
can be studied \citep{paper2}.
Hence we closely mimic observational work by projecting 
particle positions onto a grid and then analyzing this
``CCD'' image with standard IRAF routines.
Finally, we compare the appearance of our fitted ellipses to
the known conditions in our simulated satellite, with the
intention of relating observables to the satellite's
physical state.

We present our numerical and analysis methods in \S 2, 
examine the relationship between observable features in
the surface brightness distribution and the physical state
of the satellite in \S 3 and discuss how to interpret 
extra-break populations in \S 4.
We summarize our results in \S 5.

\section{Methods}

\subsection{The Simulations}

\subsubsection{Technique}

This paper is based on an analysis of five
simulations of tidal disruption (hereafter Models 1--5)
run using the technique 
described in \citet{jhb96}.  The parent galaxy 
is taken to be smooth, static and
axisymmetric, and is represented by a three-component model, 
in which the disk is described by the Miyamoto-Nagai (1975)
potential, the spheroid by a \citet{h90} model, and the halo by a
logarithmic potential:
\begin{equation}
        \Phi_{disk}=-{GM_{disk} \over
                 \sqrt{R^{2}+(a+\sqrt{z^{2}+b^{2}})^{2}}},
\end{equation}
\begin{equation}
        \Phi_{spher}=-{GM_{spher} \over r+c},
\end{equation}
\begin{equation}
        \Phi_{halo}=v_{halo}^2 \ln (r^{2}+d^{2}).
\end{equation}
We take $M_{disk}=1.0 \times 10^{11}, M_{spher}=3.4 \times 10^{10}, 
v_{halo}= 128, a=6.5, b=0.26, c=0.7$, and
$d=12.0$, where masses are in $M_{\odot}$, velocities are in km/s
and lengths are in kpc. This choice of parameters yields a nearly
flat rotation curve at $\sim 200$km/s
between 1 and 30  kpc and a disk scale height of
$0.2 $ kpc.
The radial dependence of the vertical epicyclic frequency
of the disk ($\kappa_z$) between 3 and 20  kpc 
is similar to that of an exponential
disk with a 4  kpc scale length. 
Although this model was constructed to represent the Milky Way
specifically, we anticipate our general results will be applicable to any
satellite/parent galaxy pair by using the scaling factor
$(m_{\rm sat}/M_{\rm
Gal})^{1/3}$ described in \S 1.

\begin{figure}
\begin{center}
\epsscale{1.0}
\plotone{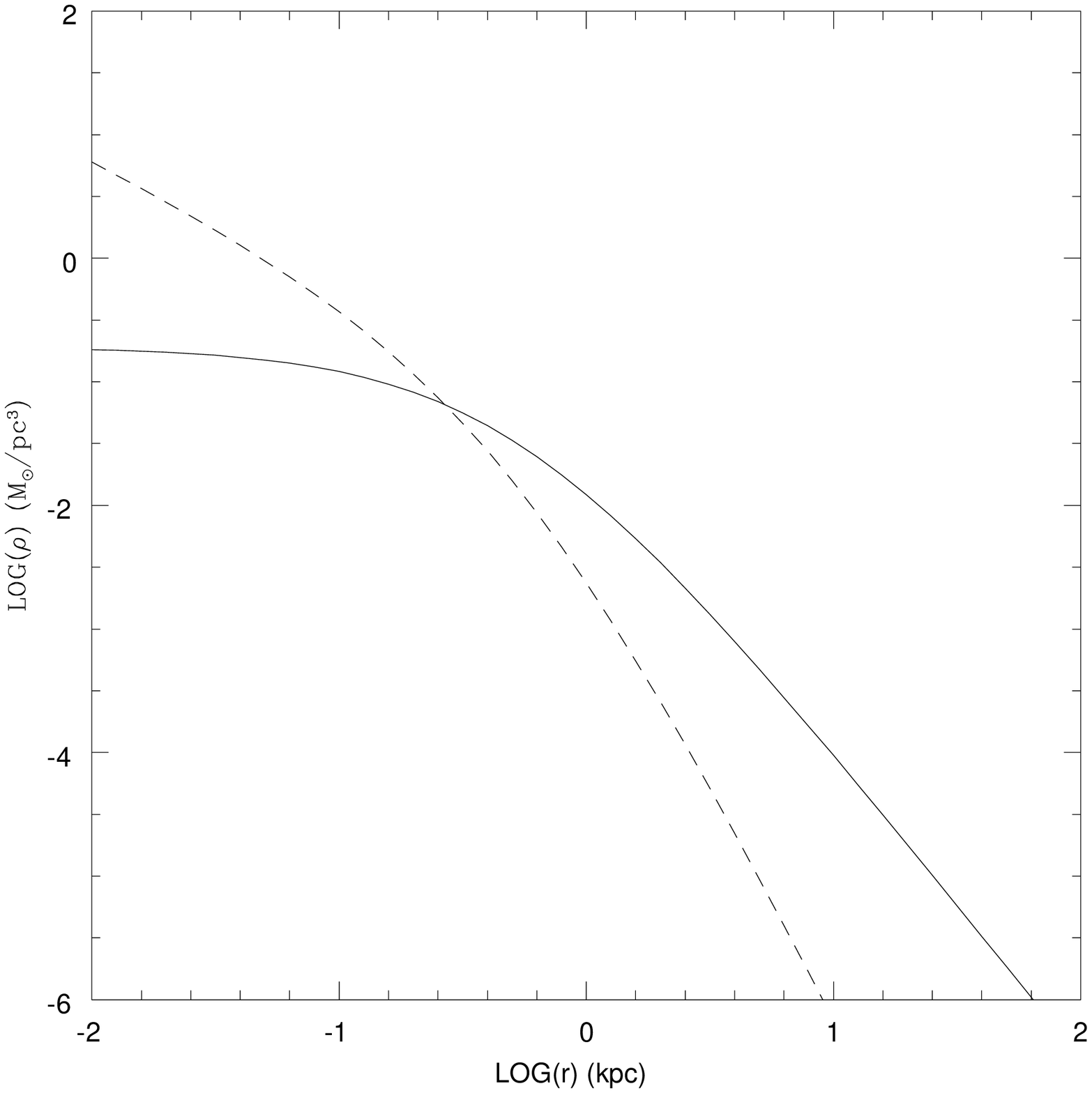}
\caption{Plummer (solid line) and Hernquist (dashed line) density profiles
of satellites used in the simulations.
\label{fig0}}
\end{center}
\end{figure}

The satellite is represented by 64,000 particles whose
mutual interactions are calculated using a code developed by
\citet{ho92} and based on the basis-function-expansion technique.
In Models 1--4 the 
satellite particles are initially distributed as a Plummer model
\citep{bt87}, and in Model 5 as a \citet{h90} model (as illustrated
in \ref{fig0}).
The total mass of the satellite in each case is
$10^8 M_{\odot}$, the scale length is 0.5\,kpc, and all are on 
orbits with pericenters at 30\,kpc. In Model 1/2/3/4/5
the satellite's orbit has an apocenter of
37/55/150/450/450\,kpc corresponding to a radial orbital time period 
(i.e. time between successive pericenters) of 
0.76/0.96/2.2/6.0/6.0\,Gyr.
In each case, the satellite is followed for a total of
5 radial orbits, with 50 snapshots saved on each orbit to allow the morphology 
to be analyzed as a function of orbital phase.
Note that each satellite maintained a core of bound particles throughout the
simulations (no satellite lost more than 60\% of its mass during the 5 orbits).

\subsubsection{Limitations of Model and Parameter Choices}
\label{param}

This study is concerned with the appearance of slowly
disrupting, low mass satellites on orbits that do not
interact significantly with any disk components of the parent
galaxy. 
We expect similar features (though a different dependence
on orbital phase) around satellites which have lost
mass through disk shocking (see Combes, Leon \& Meylan, 1999, for
a detailed discussion).

Our parent galaxy was represented by a static
potential.
We do not follow its response to the satellite and
hence cannot assess the dynamical influence of that
response on the satellite and associated debris.
We are instead interested in features in 
the vicinity of the satellite, where debris, once unbound, disperses
within an orbital time.
As long as the satellite is not massive enough for its orbit to decay
significantly within this timescale we expect our static 
potential approximation will not affect the morphology and interpretation
of such local debris.
The requirement of the orbital timescale being less than the dynamical
friction timescale corresponds to an upper limit on the mass of the satellite
(estimating $T_{\rm orb}=2\pi R/v_{\rm circ}$ 
and re-arranging equation [7-27] from
Binney \& Tremaine, 1987):
\begin{equation}
m_{\rm sat} < \left({v_{\rm circ} \over 220\, {\rm km\,s^{-1}}}\right)^2 
\left({R \over 60\, {\rm kpc}}\right)
{1 \over \ln \Lambda} 10^{11} M_\odot,
\end{equation}
where $v_{\rm circ}$ is the parent 
galaxy's circular velocity, $R$ is the distance between parent and satellite
galaxies and $\ln \Lambda$ is the Coulomb logarithm (which has typical values
of $\ln \Lambda\sim 3\>$--$\>$10 --- see Binney \& Tremaine, 1987).

We do not consider non-spherical satellites since
the force-field  a star experiences in the extra-break region 
is dominated by the monopole moment of the satellite combined with
the tidal field of the parent galaxy and is largely independent of the 
satellite's internal structure.
This argument will break down if the satellite is far from spherical
and our results should not be applied to disk systems.

We only consider non-rotating satellites.
The population of debris orbits will depend on the population of orbits 
within the satellite and this will influence the morphology in the extra-break 
region, so our results will not be directly
applicable to systems with significant
rotation.

Finally, we use single component models, which corresponds to assuming
a constant mass-to-light ratio throughout the system.
We expect that a mass-to-light ratio that decreases with distance from the
center of the satellite would change the surface brightness profile, but
not the ellipticity or orientation of the fitted ellipses. This
intuition will be tested in future work.

\subsection{Analysis}

This paper is written as a companion to the observational work
described in \citet{paper2} and the analysis performed is designed to
mimic that study as closely as possible.
In this section we describe the analysis performed on
each snapshot, and illustrate the 
the results with snapshot 121 of Model 1, viewed
at 90 degrees to the orbital plane of the satellite.

\subsubsection{Fitting isophotes}

To get `surface brightness' distributions for each snapshot the particle
positions are first projected onto the parent galaxy's rest frame in a
coordinate system centered on the satellite.  Projections with viewing
angles of 0, 30, 60 and 90 degrees to the orbital plane are used.  For a
given viewing angle, projected particle positions are binned into a
two-dimensional grid of 20\,pc$\times$20\,pc bins.  Note that although the
resulting
frames contain no conventional observational noise, they are limited by
discreteness effects due to the finite number of particles.  To overcome
this, the grids are smoothed using a 200\,pc sigma gaussian
kernel.  Ellipses are then fit to the `isophotes' (isopleths) of the
smoothed snapshots using the IRAF/STSDAS task
ELLIPSE.

\begin{figure}
\begin{center}
\epsscale{1.0}
\plotone{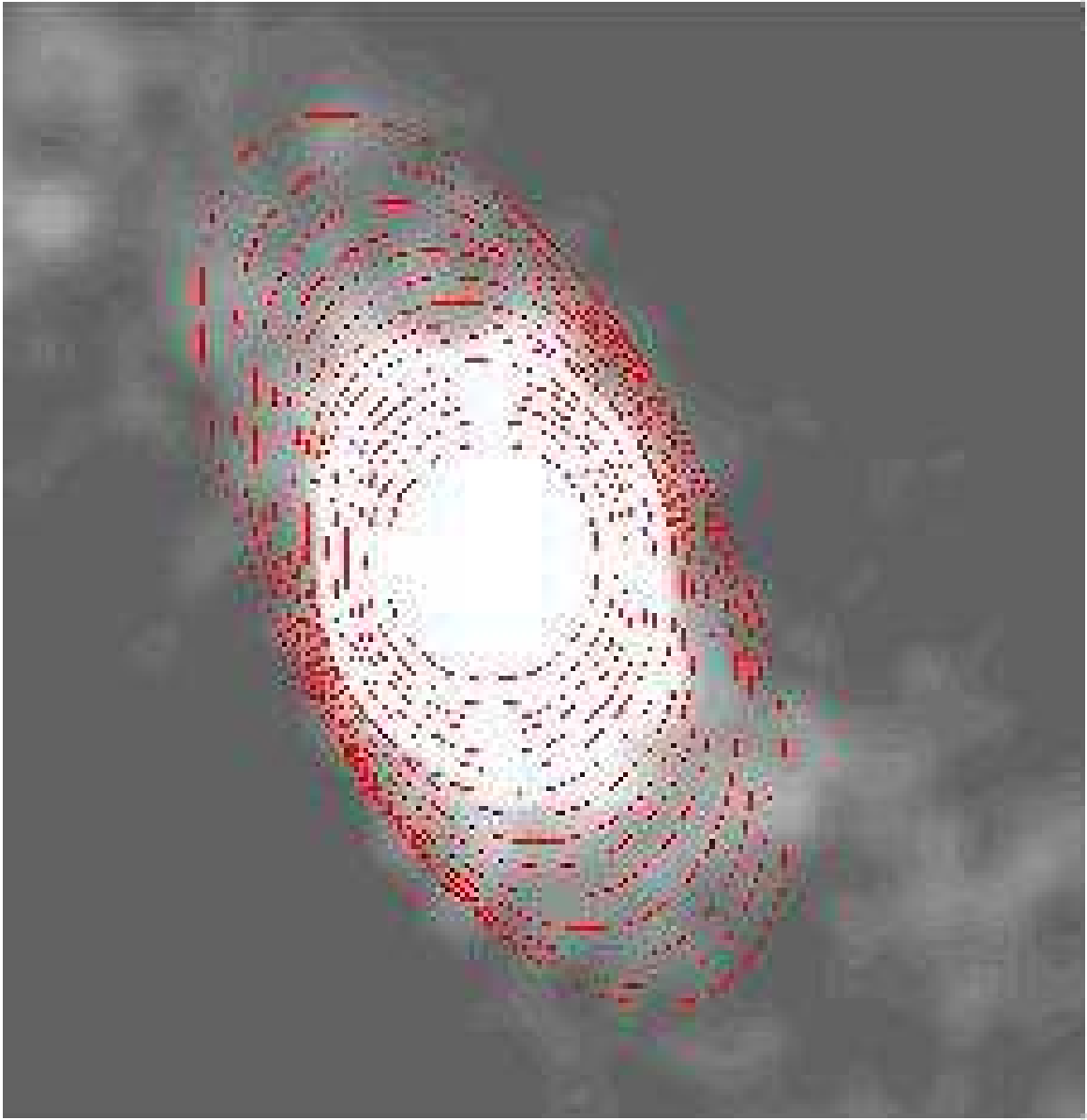}
\caption{Grayscale image of particle density 
in Snapshot 121 from Model 1, overlaid with 
results of ellipse-fitting analysis. 
\label{fig1}}
\end{center}
\end{figure}

Figure \ref{fig1} is a grayscale image of a smoothed snapshot (frame 121)
from Model 1 that has been overlaid with ellipse fits in red.

\subsubsection{Characterizing Each Snapshot}

\begin{figure}
\begin{center}
\epsscale{1.0}
\plotone{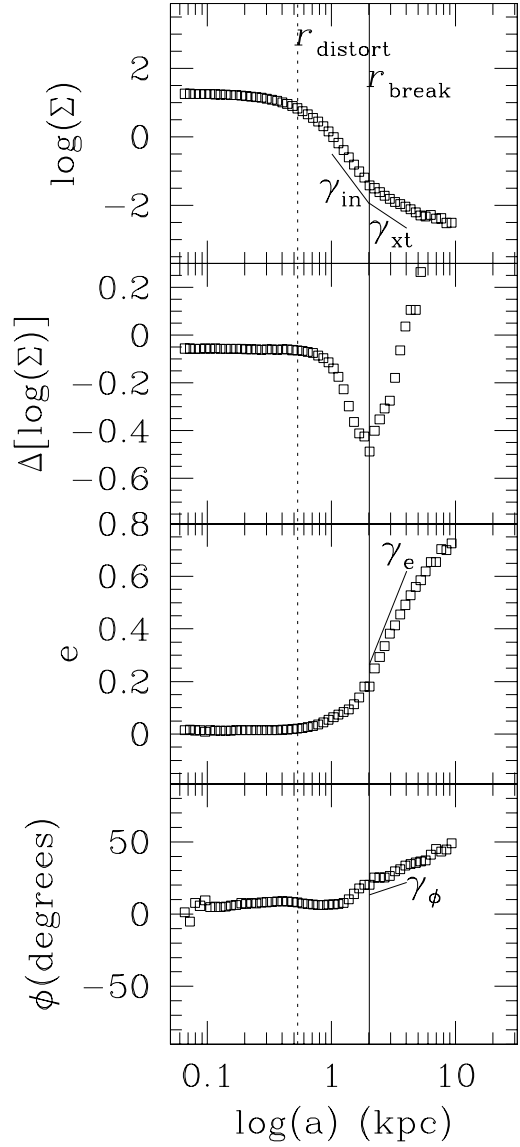}
\caption{The open squares in each panel
illustrate the results of isophotal analysis (ellipse fits)
of Snapshot 121 from
Model 1
(top to bottom): surface density (brightness) profile; difference between
instantaneous and initial brightness profiles; ellipticity; and position
angle, measured from the satellite-parent galaxy line towards the satellite's
projected direction of motion.
The dotted line running through each panel indicate
the position of
$a=\re$ and the solid line shows $a=\rb$ (see text
for explanation).
The short solid lines indicate the slopes in each quantity fitted
to points lying within a factor of two of $\rb$.
\label{fig2}}
\end{center}
\end{figure}

Given our sample of 250 snapshots from each model 
it is impossible to consider each
one separately.
Instead we define a number of quantities to summarize the 
general characteristics of each snapshot. 
As an example, Figure \ref{fig2} shows the results of the isophotal analysis 
applied to Snapshot 121 of Model 1.

The top panel of 
Figure \ref{fig2} plots the surface density $\Sigma$ 
as a function of ellipse semi-major axis $a$.
We fit lines to the seven points
(spanning nearly a factor of two in radius)
defining the profile  both interior and exterior to 
each fitted ellipse
and find the difference in the slopes of the lines 
$\Delta \gamma(a)=\gamma_{\rm xt}-\gamma_{\rm in}$.
Local maxima and minima of $\Delta \gamma$ are identified 
and $\rb$ is taken to be 
the innermost turning point with $|\Delta \gamma|>\Delta \gamma_{\rm lim}$.
We adopt a standard value of $\Delta \gamma_{\rm lim}=0.2$ ---
the dependence of results of this analysis
on the limit used will be discussed in \S 3.2.
Note that this approach means that
no breaks can be found beyond $a \sim 5\,$kpc in our 20\,kpc$\times20$\,kpc
analysis box --- we repeat the analysis on a larger
box in cases where no break is found within 5\,kpc.
The second panel emphasizes the breaks by 
plotting the difference between the current 
and the initial brightness profiles.

The lower panels of Figure \ref{fig2} show the run of ellipticity $e$
and position angle $\phi$
with semi-major axis of the fitted ellipses.
The position angle is measured relative to the line joining the satellite and
the parent galaxy center, defined to be positive towards the satellite's
projected direction of motion.
[This sign convention breaks down when the viewpoint is along the plane of
the satellite's orbit, but that is unimportant since there is no isophote
twisting in this case (see \S 3.5).]
Once $\rb$ is defined (as described in the previous
paragraph) the
rate of change of each of these quantities 
($\gamma_e$ and $\gamma_\phi$,
indicated by the solid lines in the figure) with $a$ is 
characterized by fitting straight lines to the points 
in the region $\rb< a < 2 \rb$.

Although the break in the surface density is the most striking
feature in the upper panel it is noticeable that the satellite
loses its intrinsically spherical shape well within this point.
To indicate the innermost point where the morphology might be affected by 
tidal forces we also record the innermost semi-major
axis at which $e>0.02$, hereafter referred
to as $\re$.
For intrinsically elliptical satellites $\re$ can instead be
identified as the point where the ellipticity and alignment
of isophotes start changing significantly.

\subsection{Relationship between Observables in Simulations 
Versus Real Galaxies}

By analogy with real galaxies, we are restricted to looking for profile
breaks --- it would be unrealistic, for example, to study departures from
the intrinsic/initial profile.  Real galaxies have complicated profiles
so it is easier in general to look for extreme breaks (say with
$\Delta\gamma_{\rm lim}=1.0$) than subtle ones ($\Delta\gamma_{\rm lim}=
0.2$).  While we explore the use of different $\Delta\gamma_{\rm lim}$ 
in \S 3.2 and in Figures \ref{dom} and \ref{rt}--\ref{mloss},
most of the analysis in the paper in based on 
$\Delta\gamma_{\rm lim} \sim 0.2$, 
since the resulting observables in the simulations display cleaner trends
as a function of orbital eccentricity, orbital phase, and intrinsic
profile shape.

In considering the ellipticity and position angles of the model galaxy
isophotes, it is important to note that real galaxies typically have
intrinsically non-circular isophote shapes.  This makes it more difficult
to measure ellipticity changes and isophote twists for real galaxies than
for simulated ones --- while the ellipticity is a monotonically increasing
function of radius for simulated galaxies, one should not expect the same
trend in real galaxies. 

\section{Results}

A tidal field imposed on a satellite can:
(i) heat stars within the satellite;
(ii) strip stars from the
satellite once they have been sufficiently heated;
and (iii) 
cause stripped stars to disperse away from the
satellite along its orbit.
The satellite will respond to these adjustments in its internal
phase-space distribution by attempting to relax to a new equilibrium.
All of these effects can alter the surface brightness profile
and shape of the satellite, but we expect the nature of this evolution
to depend both on the orbit and the orbital phase at which it is observed.
Along circular orbits
tidal heating, stripping and debris dispersal
are steady processes. 
For orbits that are progressively more
eccentric, a larger fraction of  heating and mass loss
occurs during a smaller fraction of
the orbit around pericenter.
The rate of debris dispersal is also largest around pericenter.
The density of debris at any time around the 
satellite is influenced in opposite senses by all these
effects and hence is hard to predict analytically.
For example, on highly eccentric orbits debris rapidly disperses away
from pericenter and congregates near apocenter, where the orbital speed is
lowest, so that the debris density is highest near apocenter despite the
predominance of mass loss around pericenter.
In contrast, the relaxation of the satellite depends only on
its internal structure and is independent of its orbit.

In the following subsections we examine how observations can be
interpreted in terms of this physical picture of heating, stripping,
debris dispersal and relaxation.

\subsection{Break Radii}

In this section we investigate how we can relate 
the appearance of breaks in the surface density profile
to the physical picture outlined above
in order to develop some intuition for the cause of 
the phase and orbit dependence of observed distortions.

\subsubsection{Morphology of Debris Around a Satellite}

\begin{figure}
\begin{center}
\epsscale{1.0}
\plotone{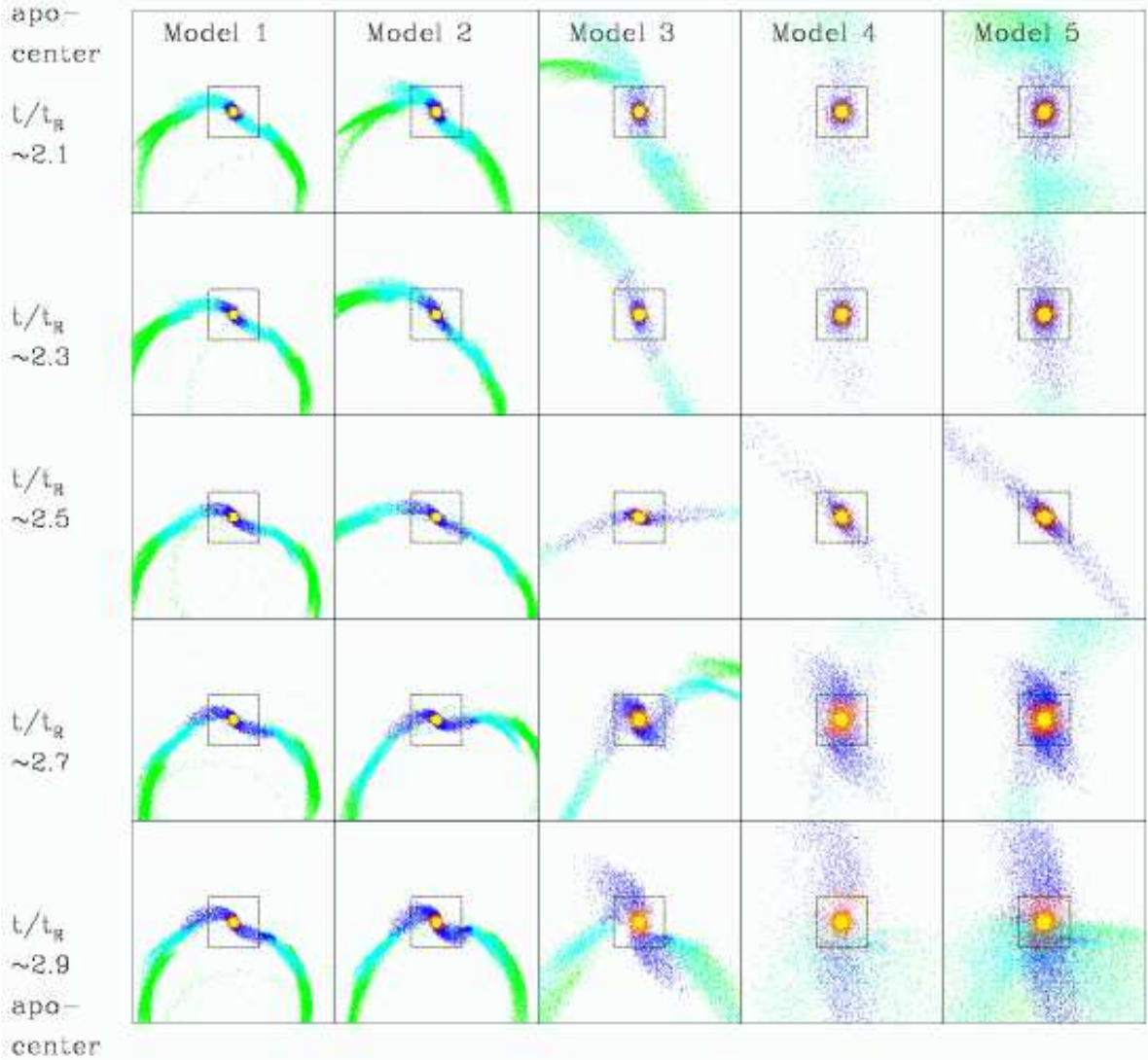}
\caption{Projection of ``debris'' particles onto the orbital plane for
snapshots equally spaced along in time along third radial
orbit of each orbit. The green/cyan/blue/red/yellow particles
are (were/will be) lost during orbits 1/2/3/4/5 according to the energy
criterion described in \S 3.1 (those not lost during the 
simulation are not shown).
The parent galaxy is on the negative $Y$-axis in all panels,
and the box size is 80\,kpc$\times80$\,kpc.
\label{snapbig}}
\end{center}
\end{figure}

\begin{figure}
\begin{center}
\epsscale{1.0}
\plotone{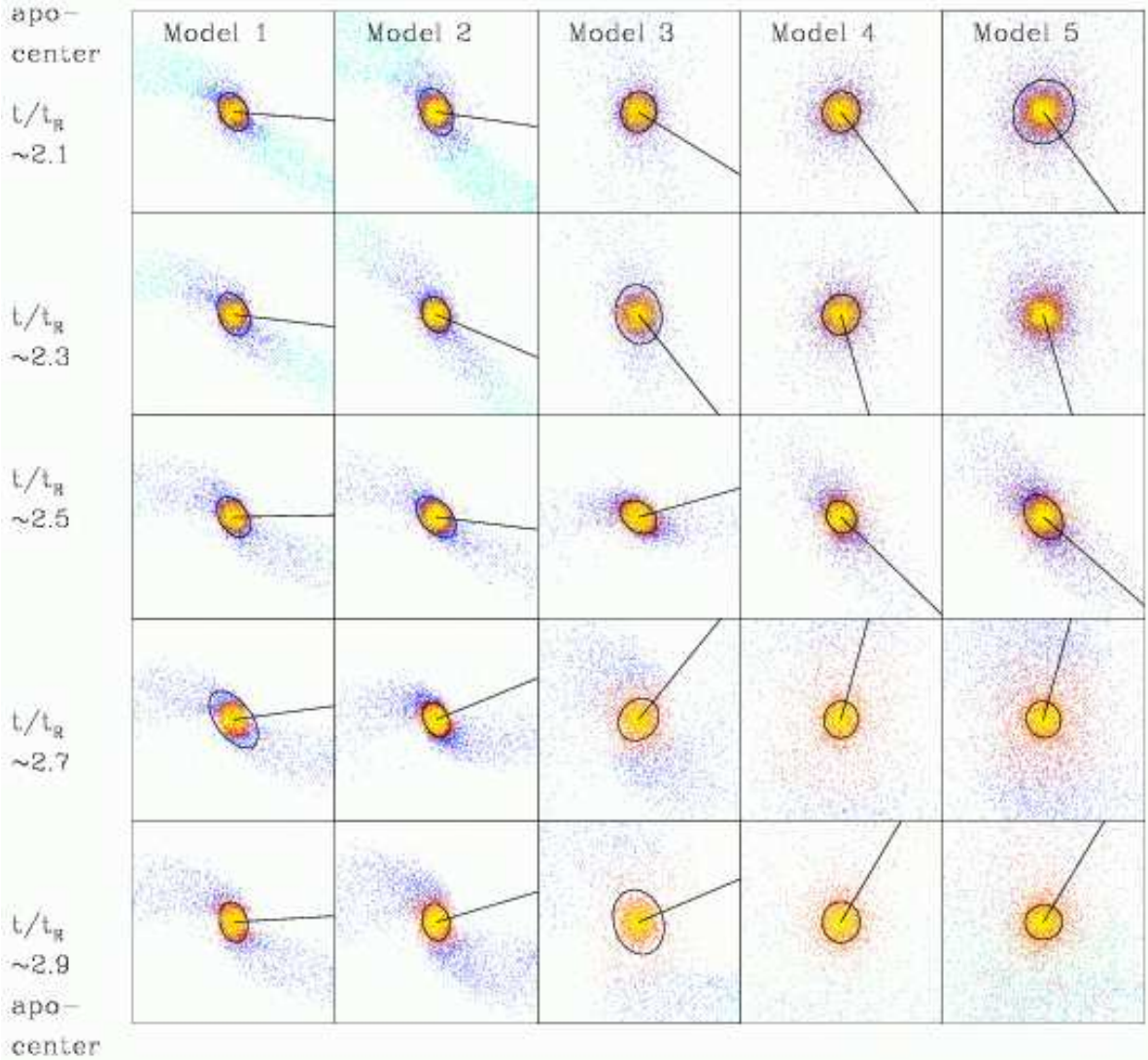}
\caption{As Figure \ref{snapbig}, but for a box size of 20\,kpc$\times$20\,kpc.
The ellipse corresponds to the isophote at $\rb$, the innermost
break radius identified using $\Delta \gamma_{\rm lim}=0.2$ in each case,
and the line indicates the velocity vector of the 
satellite. 
\label{snaplil}}
\end{center}
\end{figure}

We begin by contrasting
the morphology of 
particles that are still bound, those that are in the process of
being stripped from the satellite and those that have been lost from
the satellite for more than one orbit.
We label the particles 
with the number of the orbit (defined apocenter to apocenter)
during which they
became unbound from the satellite.
[At each time step in the simulation, we define a particle's
binding energy relative to the satellite
using the technique outlined in \citet{jsh99}:
the parent galaxy potential is ignored and the binding energy is calculated
from the potential energy of the satellite and 
kinetic energy of motion relative
to the satellite.]
Figure \ref{snapbig} illustrates the results
of this division of the particles into
separate ``debris populations'' with a set of projections of
particle positions onto the orbital plane for a series
of snapshots taken from the third radial orbit of each model
(equally spaced between the third and fourth apocenters)
The particles are coded green/cyan/blue/red/yellow
if they were (are being/will be) lost during orbits 1/2/3/4/5
(those not lost during these orbits are not shown).
In each panel, the
negative $Y$ axis points in the direction of the parent 
galaxy and
the dotted box indicates the region over which the analysis was performed.
Each panel is 80\,kpc$\times$80\,kpc.
Figure \ref{snaplil} zooms in on the 20\,kpc$\times$20\,kpc analysis boxes
indicated by the dotted lines in Figure \ref{snapbig}.
In these panels the
the ellipse indicates the isophote corresponding to  the 
the innermost break radius identified with $\Delta \gamma_{\rm lim} \sim 0.2$
(hereafter simply referred to as $\rb$)
and the line shows the direction of motion of the satellite.
No ellipse is drawn in the second panel of Model 5 since no
break with $\Delta \gamma > 0.2$
was found within the analysis box for this snapshot.

Three trends in the morphology of debris populations are worth
noting in Figures \ref{snapbig} and \ref{snaplil}.
First,  it is clear that
the extent of the particle distributions are similar
near pericenter (middle row of panels)
as all models are experiencing the same potential
field at this orbital phase.
As the distance from the center of the galaxy increases, the potential
field becomes weaker and hence the particles are most widely
distributed at the apocenters of the most eccentric orbits.
Second, although the density of unbound (green and cyan)
particles around Model 4
is always lower than Model 5, the panels appear 
otherwise morphologically similar.
Recall that these models both follow the same orbit
and differ only in their initial satellite profiles.
The difference in density reflects the lower mass loss rate from Model 4,
while the similarity in morphology is
a consequence of the fact that debris dynamics depends primarily
on the satellite mass and orbit and  parent galaxy potential and is largely
unaffected by the internal structure of a satellite
(see \S \ref{param}). 
The internal structure {\it does}
affect the identification of break radii and hence the interpretation
of extra-break populations --- this will be discussed further in \S 3.3.
Third, in Figure \ref{snapbig} the spatial
extent of each population increases steadily for Models 1 and 2 as
debris disperses along the orbit.
In contrast, although there is an increase in the extent of (for example)
the blue population from apocenter to apocenter in Models 3-5, the
particles actually spread further along the orbit
at pericenter than they do in the
final apocenter panels.
This difference is a direct consequence of the variable rate of debris
dispersal with orbital phase along the more eccentric orbits.
Similarly, \citet{hw99} found peaks in the density at the center
of the remnant of a destroyed satellite at both pericenter and apocenter as 
well as a general trend for the density to decrease with time.
These peaks in density are not so obvious in our plots because of the scale: 
within each panel we are looking at debris spread around a bound satellite
along a range of orbital phases.

\subsubsection{Appearance and Physical Interpretation of Break Radii}

Figure \ref{prof} shows the surface density profiles corresponding to
each of the panels in Figure \ref{snapbig}, with the solid lines indicating
$\rb$. The stars in the panels for Models 3--5 plot the results of
the same analysis run on a 40\,kpc$\times$40\,kpc box.
Figure \ref{dprof} plots the difference between the current
profile and the initial profile at the start of the simulation in order 
to emphasize the validity of the identification of the
shallowest breaks that are not apparent to the eye in Figure \ref{prof}.
It also demonstrates 
that the profile has evolved significantly away from
the initial profile even at radii well
within the innermost identified break radius.
In particular note that the central surface density decreases ever so
slightly but steadily
throughout each simulation --- a signature of the expansion of the satellite
as it relaxes in response to the heating and stripping of its outer regions.

Note also that there are several cases of multiple breaks appearing in 
a panel. Such features have also been found in simulations of disk shocking
of globular clusters \citep{lmc00}.

Figures \ref{snapbig} and \ref{snaplil} both indicate that
although the different debris populations overlap,
there are clear boundaries between regions dominated by each set,
and these provide a useful backdrop against which to
compare $\rb$.
[This clear separation between particles lost on each pericentric
passage is analogous to \citet{clm99}
finding that each disk passage
in their simulations of globular cluster disruption produced
distinct overdense ``lumps'' in their tidal tails.]
The left hand panels of
Figure \ref{dom} assesses 
which population dominates the surface brightness profile around $\rb$
by plotting
the number of particles in each debris population (color coded in the
same way as Figure \ref{snapbig}) in the region where the fitted
ellipses have semi-major axes $0.9\rb<a<1.1\rb$.
The aim of this figure is to understand whether breaks can be attributed to:
{\it heating} within the still-bound population; the {\it stripping} of
particles from the satellite; or
the {\it transition} between two unbound populations lost on different
(consecutive) orbits.
Each row of the figure shows the same sequence of colors dominating the
particle numbers as each population is heated and stripped in order.
However, both the phase at which the different colors become dominant
and the ratio of the number of particles
in other populations to the dominant one at any given phase are orbit
dependent.
This variance suggests that there is no unique physical interpretation
of breaks in a surface density profile.
In the next section, we go on to examine the orbit and phase  
dependence of break interpretation in more detail.

\subsection{Dependence on Orbital Eccentricity and Phase}

In this section we first examine how the nature of the break in the surface
brightness profile at $\rb$ depends on the
orbit of the satellite. 
We then investigate whether the ellipticity and position
angles of the isophotes might allow us to distinguish between breaks caused
by different processes.
Finally, we look at how general our results are by summarizing
the observables as a function of orbital phase for all snapshots from
each model.
We concentrate exclusively on Models 1--4, which all had the
same initial satellite profile. In \S 3.3 we will go on to
examine the dependence on satellite profile by
comparing Models 4 and 5. 

\subsubsection{Orbital Eccentricity}

\begin{figure}
\begin{center}
\plotone{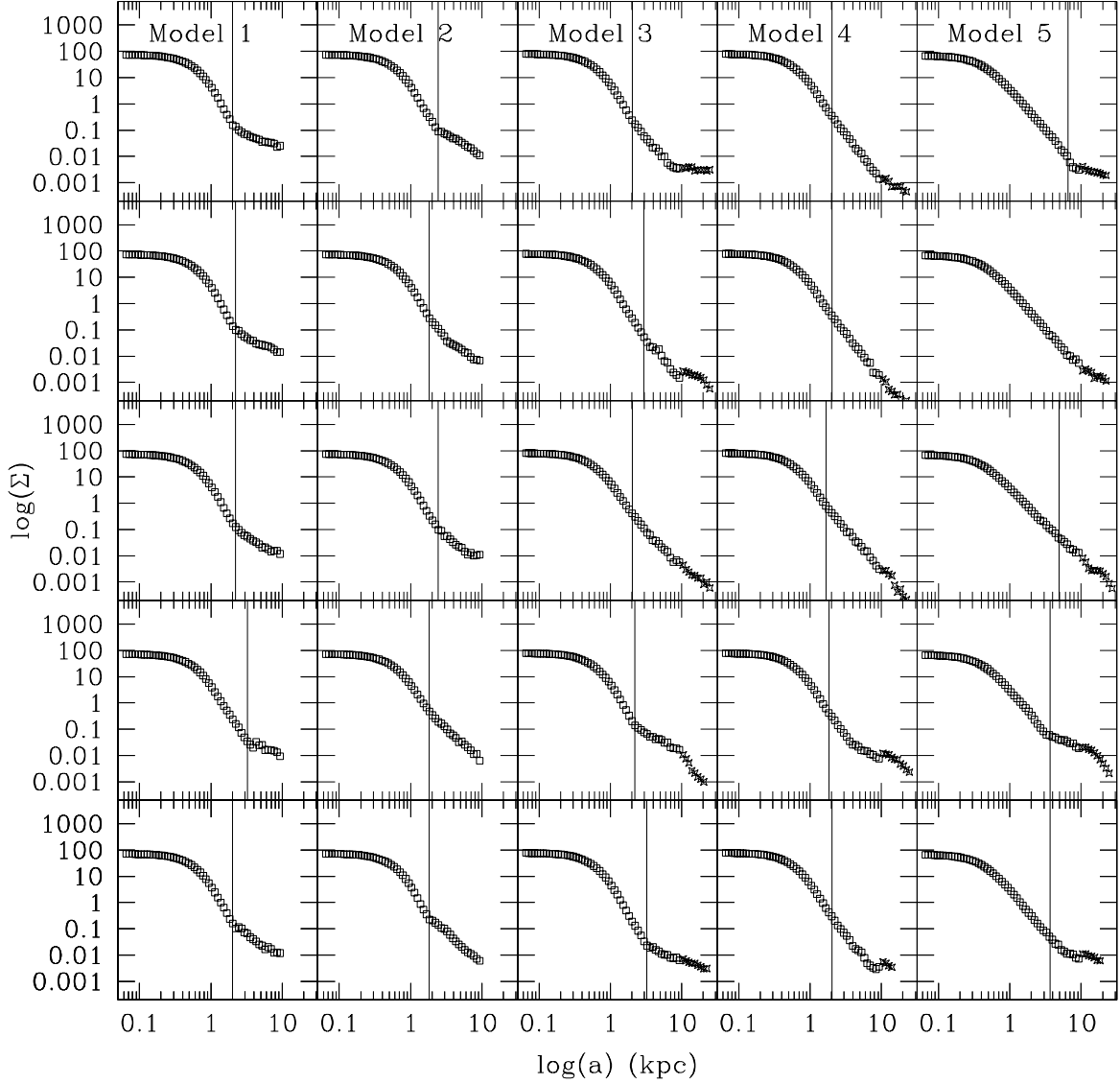}
\caption{Surface density profiles corresponding to each of the
panels in Figure \ref{snaplil}. The open squares are from the analysis
on the 20kpc$\times$20kpc box and the stars are for the analysis on the 
40kpc$\times$40kpc box
The solid line indicates the
location of $\rb$. No line is included in the second panel of Model 5
because no break was found in this case.
\label{prof}}
\end{center}
\end{figure}

\begin{figure}
\begin{center}
\plotone{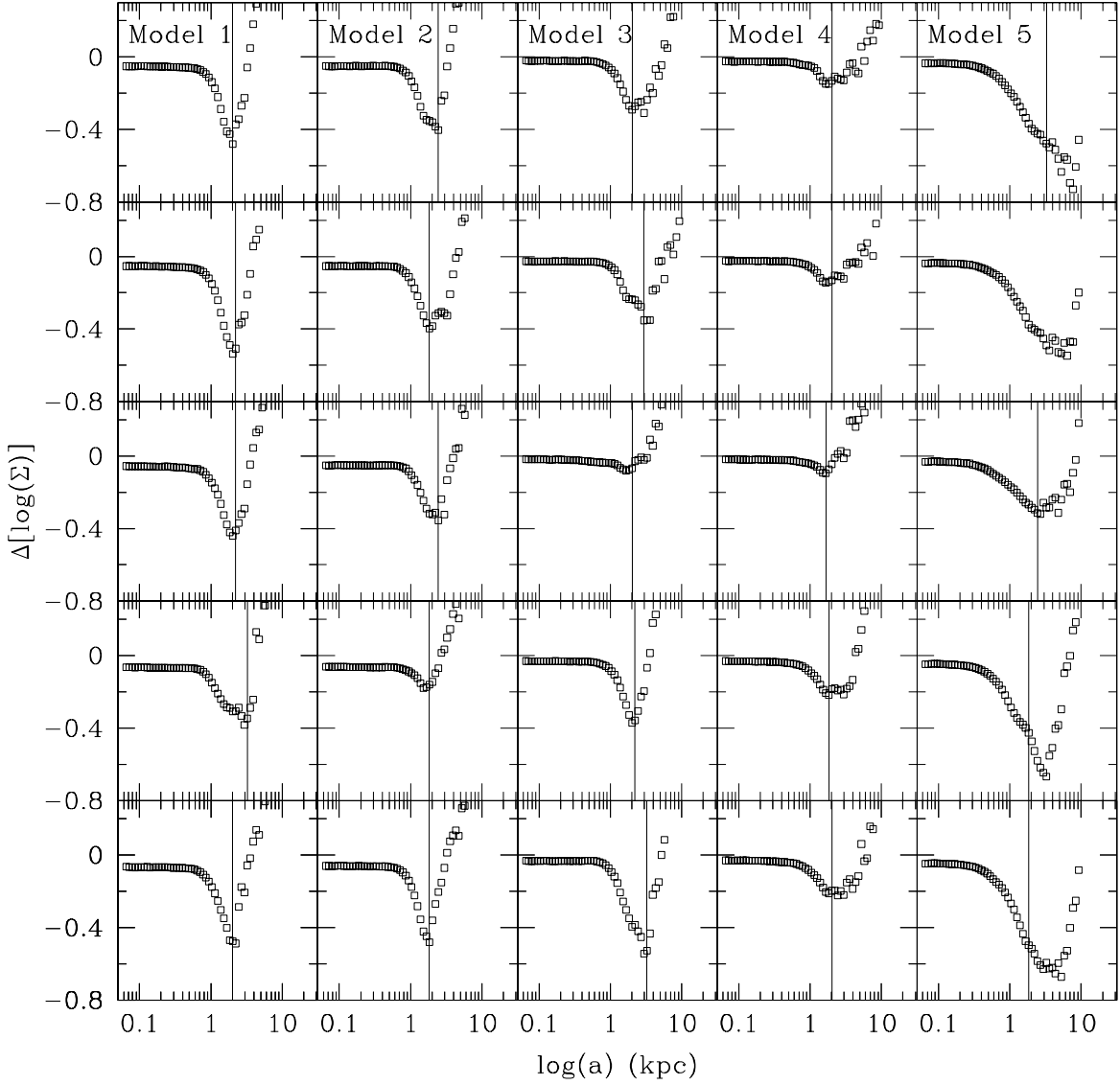}
\caption{The difference between the current and inital surface density profile
of the satellite corresponding to each of the
panels in Figure \ref{snaplil}. The solid line indicates the
location of $\rb$.
\label{dprof}}
\end{center}
\end{figure}

\begin{figure}
\begin{center}
\epsscale{1.0}
\plotone{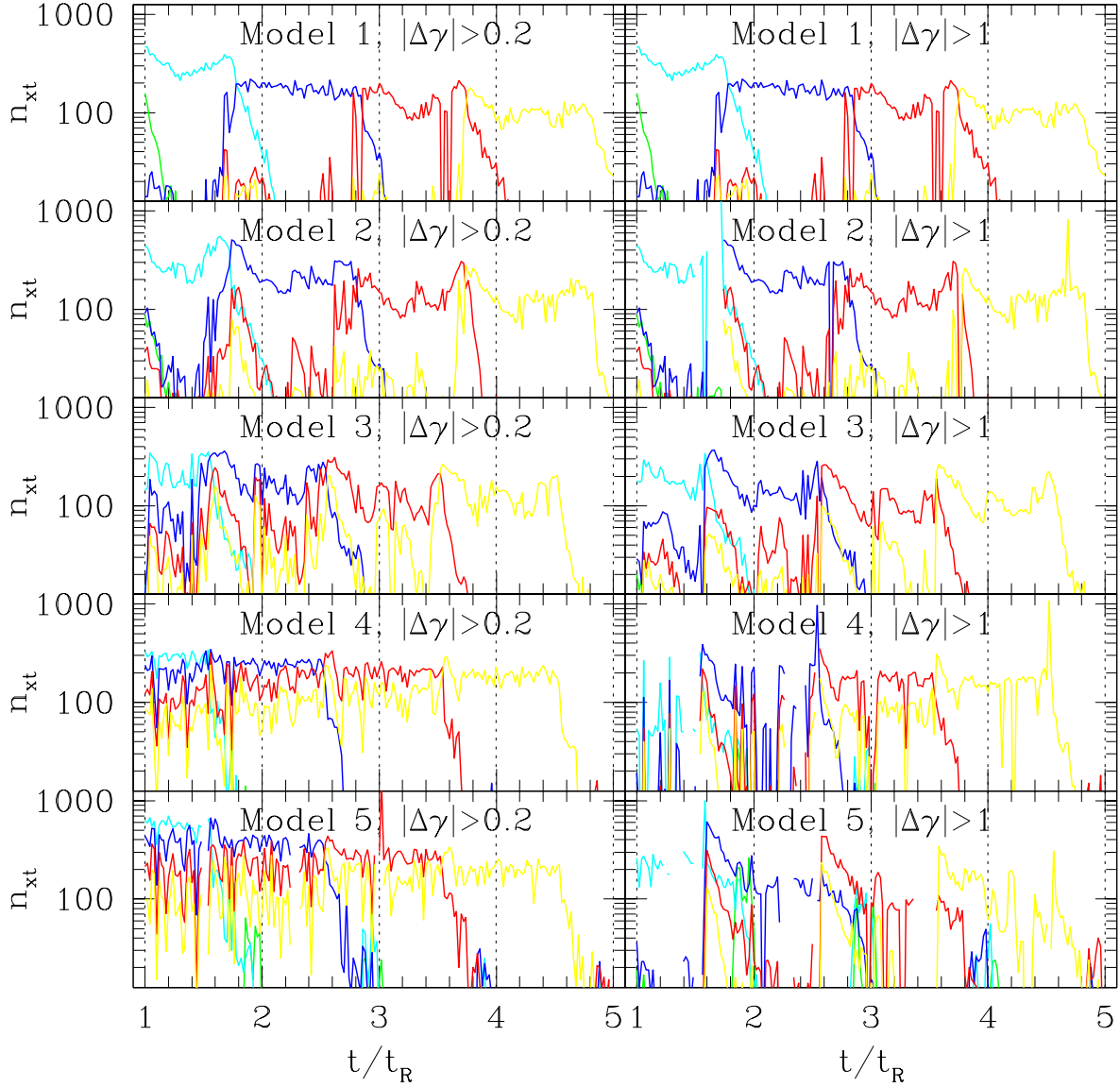}
\caption{The number of particles found in the region $0.9 \rb < a < 1.1 \rb$
from each of the debris populations (color coded as Figures \ref{snapbig} 
and \ref{snaplil}) for $\Delta \gamma_{\rm lim}=0.2$ (left hand panels)
and $\Delta \gamma_{\rm lim}=1.0$ (right hand panels).
\label{dom}}
\end{center}
\end{figure}

Along near-circular orbits (e.g. Model 1) all panels of
Figure \ref{prof} show clear breaks above the inner profile, with
the extra-break population having similar 
amplitude and a shallow slope at all phases.
Figure \ref{dom} shows that 
our analysis of the extra-break population 
is dominated by debris lost on the current orbit (blue in Figures
\ref{snapbig} and \ref{snaplil}) throughout each orbit, with
negligible contribution from other populations and 
sharp transitions between the dominant population just 
before the apocenter of each orbit.
These characteristics indicate
that the breaks we have identified generally occur where stripping
is actively taking place along circular orbits ---
heating within the bound population is steady and hence does not
cause an identifiable signature in the profile, and the sudden
change between bound and dispersing populations causes an
abrupt break above the inner profile.

As the orbit becomes more eccentric the extra-break population has
varying amplitude and slope, but in general a higher amplitude and steeper 
slope than the  circular case.
Figure~\ref{dom}
shows that the transition between the different dominant populations
identified around $\rb$
moves to {\it earlier\/} orbital phases for more eccentric orbits.
Moreover, the {\it other\/} (i.e. non-dominant) populations make
an increasing fractional contribution throughout every orbit for
increasing orbital eccentricity.
In the most extreme cases (e.g. Model 4) the contribution of particles
to be lost on the next orbit is of the same order as that
of particles currently being stripped.
Finally, since most mass loss occurs
at pericenter along these eccentric orbits the vast majority of
particles labeled as being lost on the current orbit (e.g. the
blue particles which dominate for $2<t/t_R<2.5$) will still be bound
during the first half of the orbit; in the second half of each orbit the
dominant population consists of particles to be lost during the next orbit
(e.g. the red particles for $2.5<t/t_R<3$). 
These characteristics suggest that the $\rb$ identified lies
well within the still-bound population as a result of shock-heating
during the pericentric passages.

Note that additional breaks in the profile beyond
$\rb$ occur in Model 4, at larger radii
and with larger $|\Delta \gamma|$ than those at $\rb$
(see Figures~\ref{prof} and \ref{dprof}).
Indeed, if we make our condition for break identification more restrictive
($\Delta \gamma_{\rm lim}=1.0$), the break identified is often this outer,
more-apparent break.
The relative dominance of a single (unbound) debris
population, rather than a mixture of bound populations, at most
orbital phases in the
right hand panels of Figure \ref{dom} indicate that this outer,
more extreme break is more analogous to the breaks caused by stripping
as seen in near-circular orbits.

A further example of multiple breaks within a single profile
having different interpretation is the fourth panel of Model 3
in Figure~\ref{prof}.
Here, $\rb$ appears to
correspond to a point within the satellite that has just been
shocked by the pericentric passage. The clear break below the continuation
of the surface density at about $a=10$\,kpc, on the other hand, can be
attributed to the morphological differences between the
blue and cyan particles (apparent in Figures
\ref{snapbig} and \ref{snaplil}), or the transition 
between two unbound populations.

We conclude that breaks in the profiles of objects disrupted along
near-circular orbits will usually be due to stripping.
Objects on more eccentric orbits are more severely
shocked at pericentric passages and can exhibit breaks within
their still-bound populations as well as breaks analogous to those seen
along near-circular orbits.
Unfortunately, the orbit of a satellite is not an observable quantity
so the exact interpretation of a break in a profile is ambiguous without
some other clues.

\subsubsection{Relating the Nature of $\rb$ to Other Properties}

\begin{figure}
\begin{center}
\epsscale{1.0}
\plotone{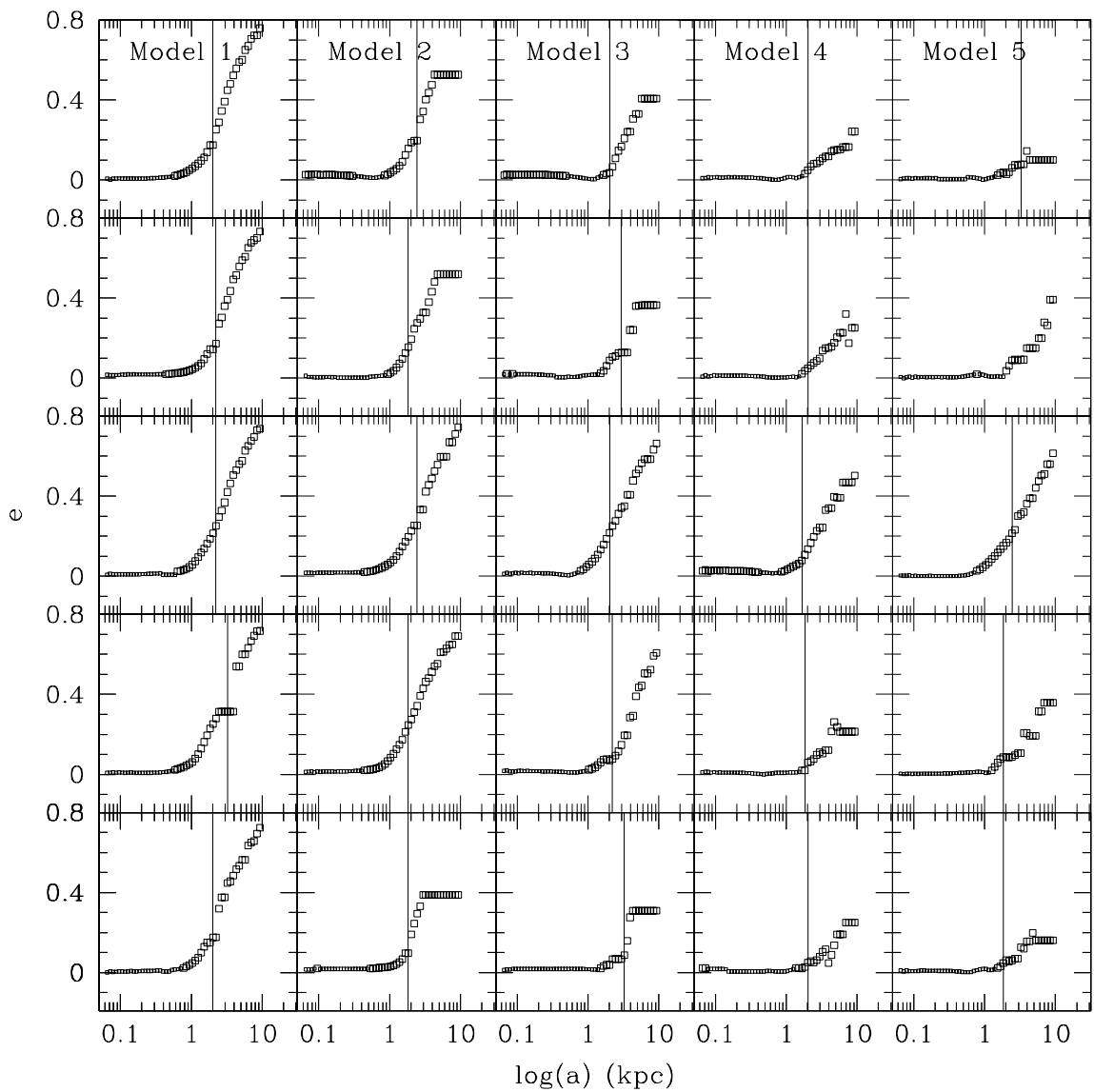}
\caption{Ellipticity profiles corresponding to each of the
panels in Figure \ref{snaplil}. Miniature squares are plotted if 
$e<0.02$. The solid line indicates the
location of $\rb$.
\label{ellip}}
\end{center}
\end{figure}

\begin{figure}
\begin{center}
\plotone{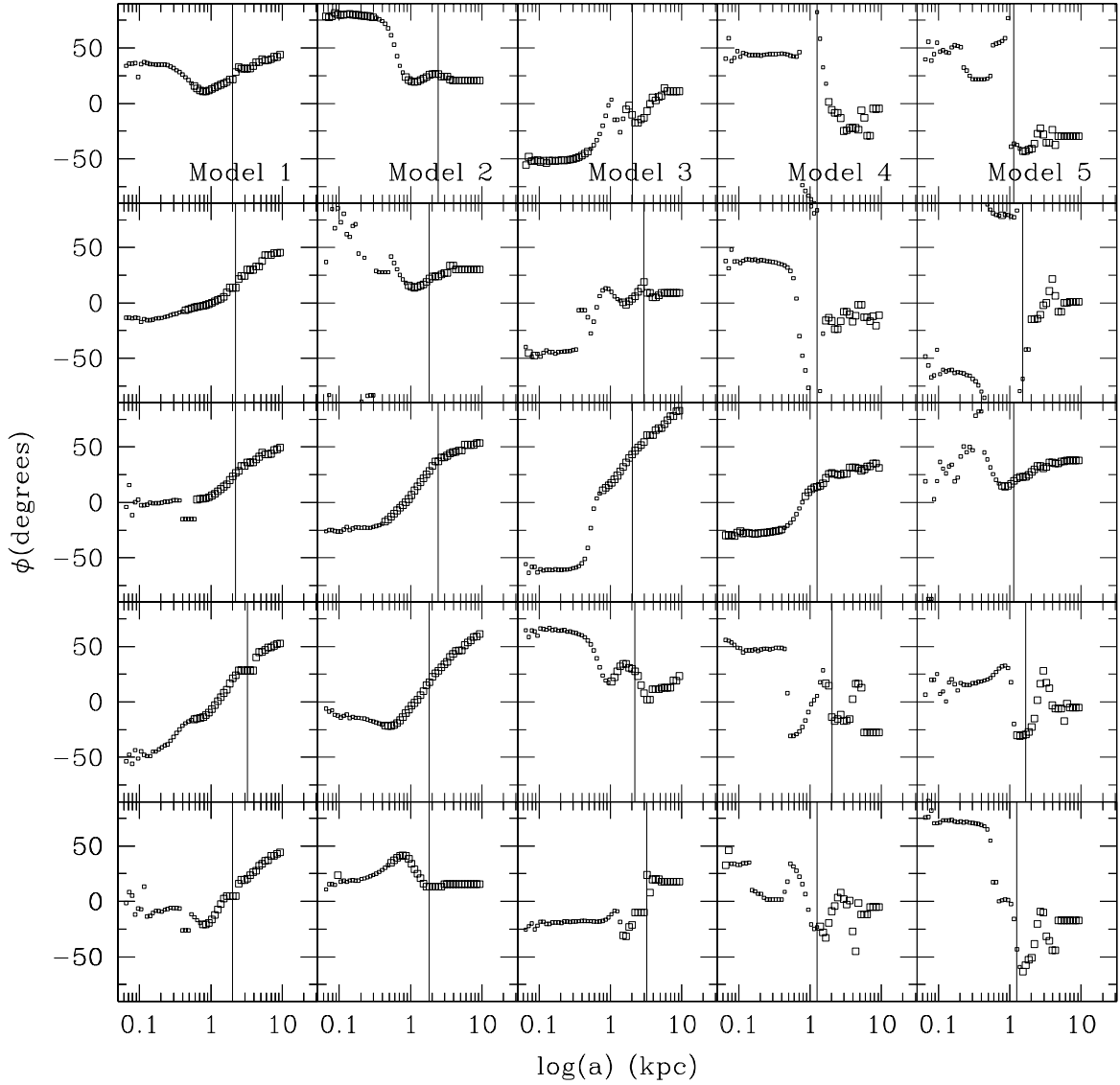}
\caption{Position angle profiles corresponding to each of the
panels in Figure \ref{snaplil}. Mini-squares are plotted if 
$e<0.02$. The solid line indicates the location of $\rb$.
\label{angle}}
\end{center}
\end{figure}

Additional information can be gleaned from the shape and orientation
of the isophotes.
Figures \ref{ellip} and \ref{angle}
show the run of ellipticity $e$ and position angle $\phi$
(with half-sized points showing 
isophotes with $e<0.02$ --- i.e. $r < \re$)
corresponding to
each of the panels in Figures \ref{snapbig}--\ref{dprof}.
The solid lines indicate
$\rb$, which
often (although not always) coincides with
features in these plots.
In general, $e$ monotonically increases with $a$
but $\phi$ does not, suggesting that
reversals in the sense of twists in tidal features may be
common.
Comparing these figures with our earlier interpretation, we find that
if $\rb$ is caused by stripping, then significant 
ellipticity and twisting of isophotes can occur well within this
point (or, equivalently $\rb/\re >1$ and the full-sized points
begin well within the break radius in Figures~\ref{ellip} and \ref{angle}).
If $\rb$ is caused by heating it more closely corresponds to
$\re$.
In the latter case we can also conclude that the satellite is likely to
be on an eccentric orbit.

\subsubsection{Trends in Observables with Orbital Phase}

\begin{figure}
\begin{center}
\epsscale{1.0}
\plotone{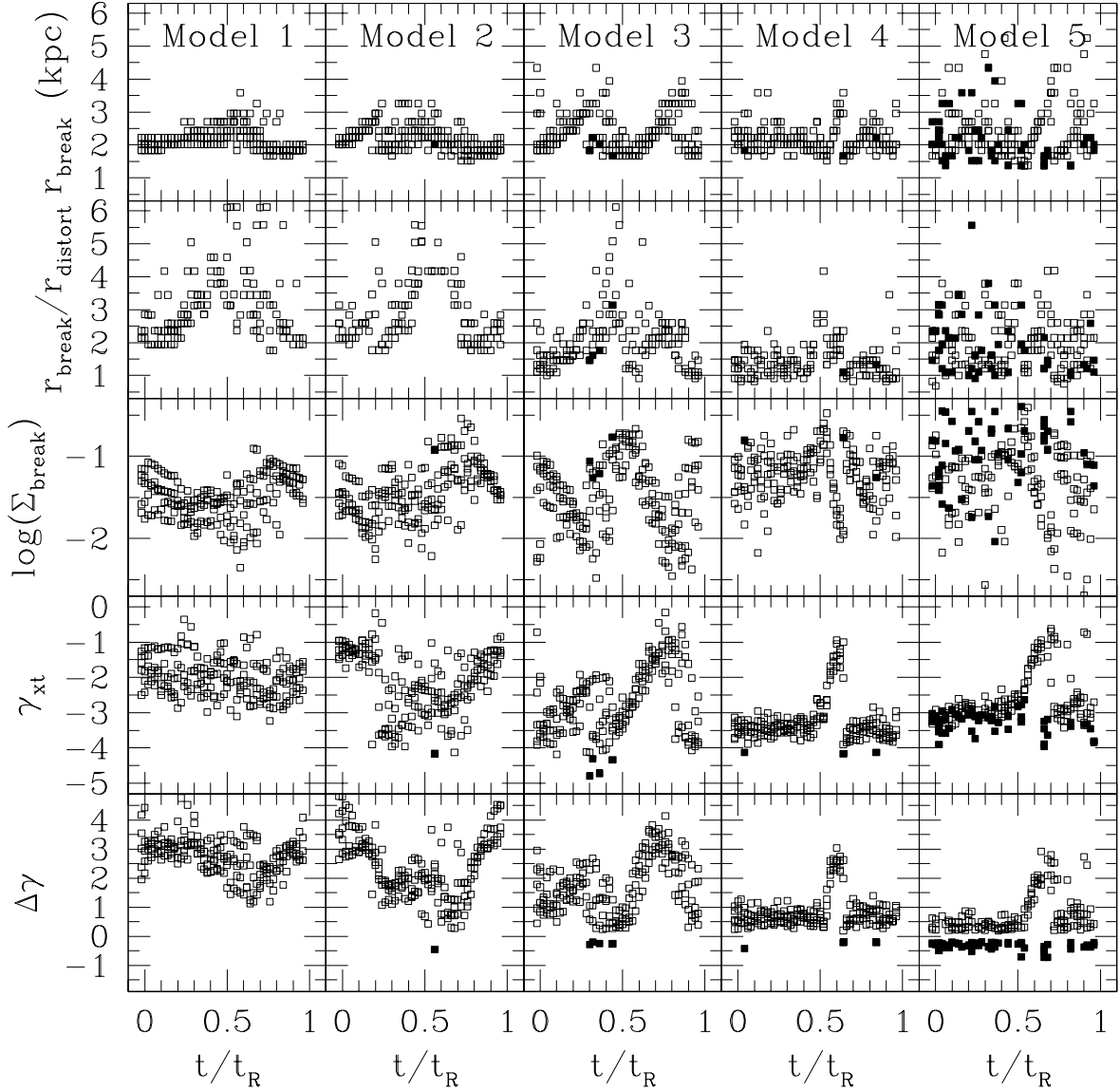}
\caption{Summary of ``observed'' quantities derived from
the surface density profiles of
all 250 snapshots of each model as a function of orbital phase
(top to bottom): radius at which the first significant break occurs in the
brightness profile; ratio of $\rb$ to the radius at which the isophotes start
becoming significantly distorted; surface brightness at $\rb$; slope of
brightness profile beyond $\rb$; and change in slope of the brightness
profile at $\rb$.
Apocenter is at $t/t_R=0$ and 1, pericenter is at $t/t_R=0.5$.
The solid squares highlight points where $\Delta \gamma < 0$.
\label{profobs}}
\end{center}
\end{figure}

\begin{figure}
\begin{center}
\epsscale{1.0}
\plotone{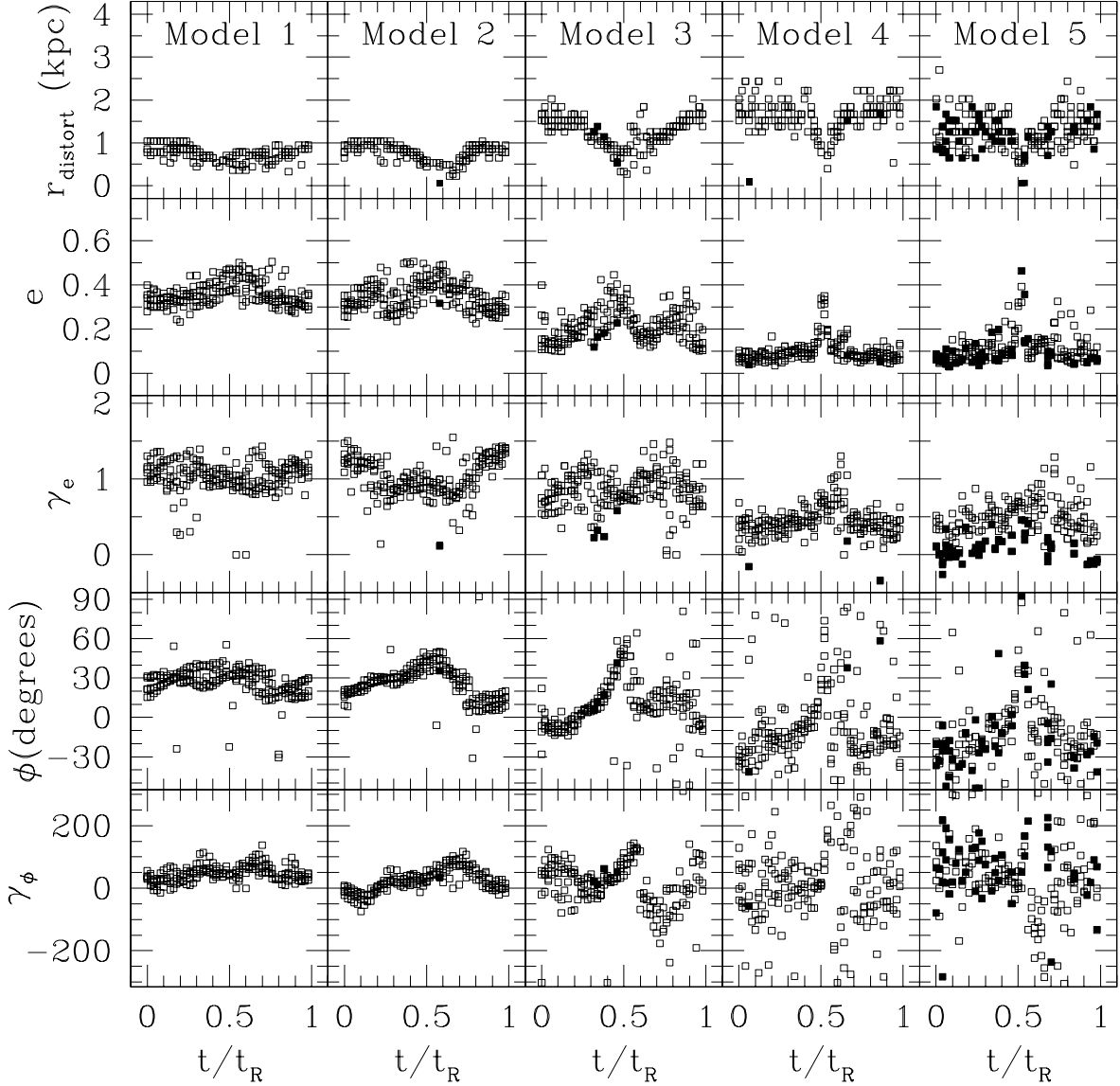}
\caption{Summary of ``observed'' quantities from the isophotal shapes of
all 250 snapshots of each model as a function of orbital phase
(top to bottom): radius at which the isophotes start becoming significantly
distorted; ellipticity at $\rb$; rate of change of ellipticity with radius
beyond $\rb$; position angle at $\rb$; and rate of change of position angle
with radius beyond $\rb$.
Apocenter is at $t/t_R=0$ and 1, pericenter is at $t/t_R=0.5$.
The solid squares highlight points where $\Delta \gamma < 0$.
\label{ellobs}}
\end{center}
\end{figure}

We examine how general the conclusions in the previous sections
are by summarizing the ``observations'' (using quantities defined in
\S 2.2.2) of all the snapshots 
in the simulations.  These are
presented in Figures \ref{profobs} and \ref{ellobs}, again 
viewed along a line of sight perpendicular
to the orbital plane.
There are clear trends in all 
quantities with orbital phase, and a large
scatter about these trends.
The nature of the trends is orbit dependent, with small-amplitude,
continuous changes in the case of steady tidal stripping along
the near-circular orbit (Model 1) and sudden swings over
a progressively smaller fraction of the orbit as the orbital
eccentricity increases.

One striking feature in Figure~\ref{profobs}
is the increase in $\gamma_{\rm xt}$ along the most eccentric orbit
(Models 4 and 5) 
after the pericenter shock followed by the sudden dive to
a new $\gamma_{\rm xt}$ which remains steady throughout the
rest of the orbit --- this occurs at the orbital
phase clearly corresponding to the transition between dominance by
different populations, shown in Figure \ref{dom}.
One interpretation is that the increase is caused by $\rb$ first tracking 
the population that has just been unbound at pericenter 
(so the break at this orbital phase
is analogous to the breaks caused by particles being stripped
in Models 1 and 2) and then readjusting as the 
disturbance in the inner profile caused by heating within the
bound satellite becomes apparent.

This idea is supported by the general $e$ and $\phi$ characteristics
seen in Figure~\ref{ellobs}.
Comparing across the orbit types,
we find $e$, $\gamma_e$, $\phi$ and $\gamma_\phi$ are
systematically lower at most phases along the more eccentric orbits
since $\rb$ was identified
well within the bound population of the satellite.
The only phase where this is not true corresponds to where
we expect stripping to be occurring around $\rb$.
These characteristics are consistent with our findings in the previous
section that breaks in the brightness profile
caused by heating tend to occur close to the
point where the satellite first becomes deformed by the
tidal field (i.e. where $\rb/\re \sim 1$ --- see second row of panels
in Figure ~\ref{profobs}), while those caused by stripping
can be well outside this region.
Thus, our use of spherical satellite models would allow
us to attribute the breaks to
heating or stripping "observationally" using $\rb/\re$,
since $\re$ always occurs within populations still
bound to the satellite.
Unfortunately, $\re$ may not be so unambiguously defined in 
intrinsically elliptical systems, although a similar aim might be fulfilled 
by looking at the onset of isophote twists (as discussed in \S 2.2.2).

These plots confirm
that though the appearance of
breaks is both orbit and orbital phase-dependent, the 
ellipticity and twisting of isophotes can offer further clues
to a physical interpretation.

\subsection{Dependence on Intrinsic Profile Shape}

In Model 5 we repeated a simulation with the same orbit
as Model 4 but with a Hernquist rather than a Plummer model satellite.
Comparing the relevant panels in Figures \ref{snapbig}--\ref{ellobs}
we find that the characteristics of the debris in both
populations are very similar --- a reflection of the very similar dynamical
influences that particles in both models experience once they are
unbound from the satellite. Key differences are that the surface density
of the extra-break population is greater
and $\Delta \gamma$
is systematically lower throughout much of Model 5.
The former is because the mass loss rate is higher from the satellite
in Model 5 --- disruption is a density criterion so altering the
mass distribution does in general alter the mass loss rate.
The latter is because the surface brightness profile
of the Hernquist model is shallower than that of the Plummer model,
so that breaks are less pronounced.
As a consequence, along Model 5 there are instances where 
no $\rb$ is identified
within the 20\,kpc$\times$20\,kpc grid on which the analysis is
performed, even though we expect the debris in this model
to behave similarly to that in Model 4.
There are also more cases where the break is {\it below}
rather than above the continuation of the inner surface density ---
these are highlighted with solid boxes in all panels of 
Figures \ref{profobs} and \ref{ellobs}.
These breaks tend to occur well inside the 
bound satellite population in regions where the satellite is less
distorted by the tidal field (i.e. low $e$ and twists).

The comparison of the two models emphasizes that $\Delta \gamma$ is
a reflection of the initial satellite profile as much as the orbit
which the satellite is on. A more reliable indicator of orbital
eccentricity is $\gamma_{\rm xt}$ which is systematically lower
along the more eccentric orbits, independent of satellite model
(Figure~\ref{profobs}).

\subsection{Dependence on Resolution}

Near apocenter, Models 3--5 covered significantly larger
spatial scales than Models 1--2.  In order to examine
the outskirts of these models we
repeated our analyses within a 40\,kpc$\times$40\,kpc
box, effectively degrading our pixel scale by a factor of two
(results shown as stars in Figure \ref{prof}).
We found that the ``observables'' on this larger grid were
similar to their counterparts on the nominal grid (latter shown in
Figures \ref{profobs} and \ref{ellobs}).
However, the larger grid scale did smooth over the smallest
breaks so that (for example) none of the $\Delta \gamma < 0$
points in Model 5 were apparent. This gives some indication
of the observational problems in identifying these features
in data on real galaxies.  We have confirmed the robustness of the satellite profiles against
numerical noise by running a test model with 4X the number of
particles (256,000).  The resulting profiles and feature measurements
show no systematic differences from the 64,000 particle runs.

\subsection{Dependence on Viewing Angle}

\begin{figure}
\begin{center}
\epsscale{1.0}
\plotone{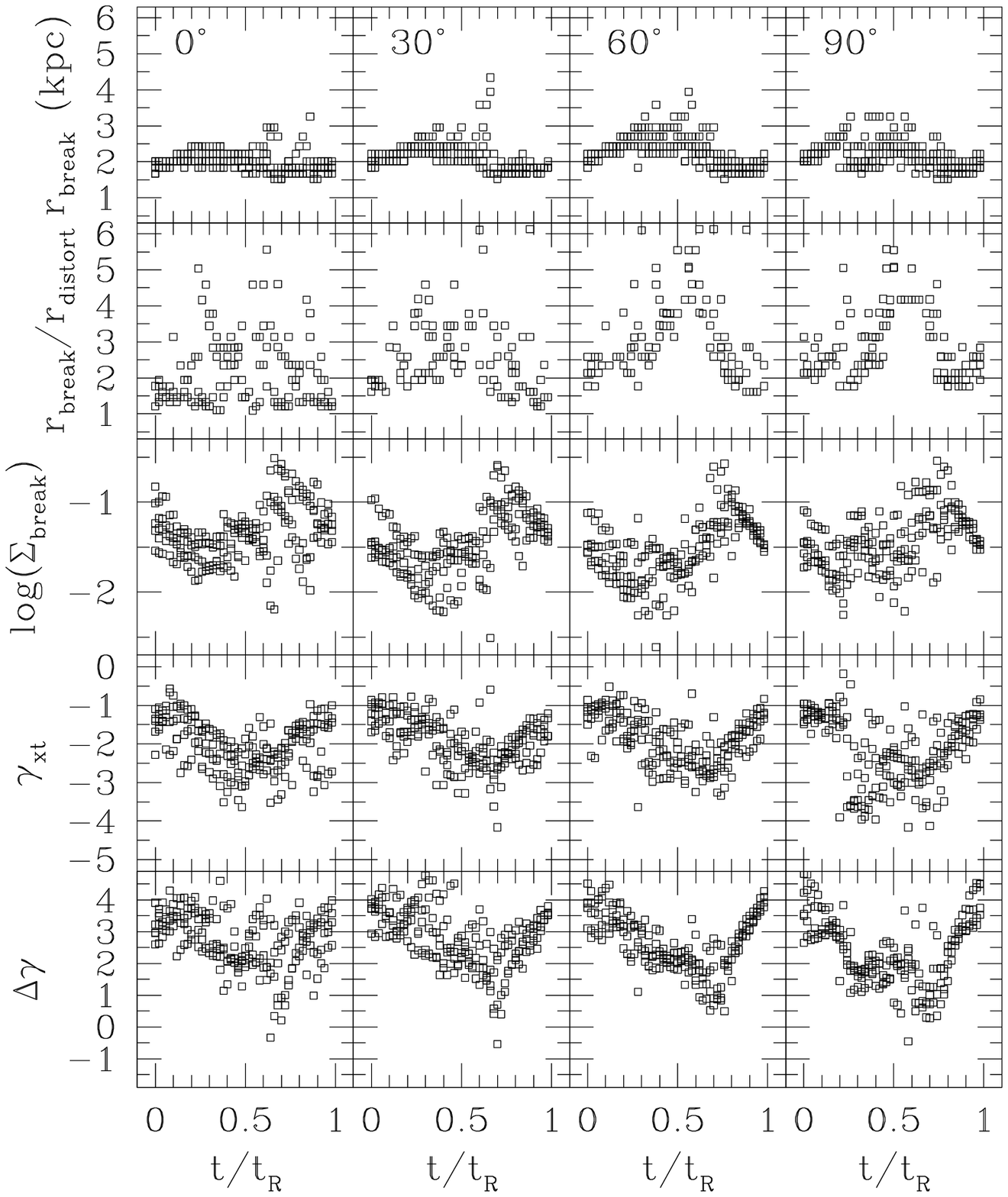}
\caption{Repeat of Figure \ref{profobs} for Model 2 viewed at
0, 30, 60 and 90 degrees to the orbital plane.
\label{pinc}}
\end{center}
\end{figure}

\begin{figure}
\begin{center}
\epsscale{1.0}
\plotone{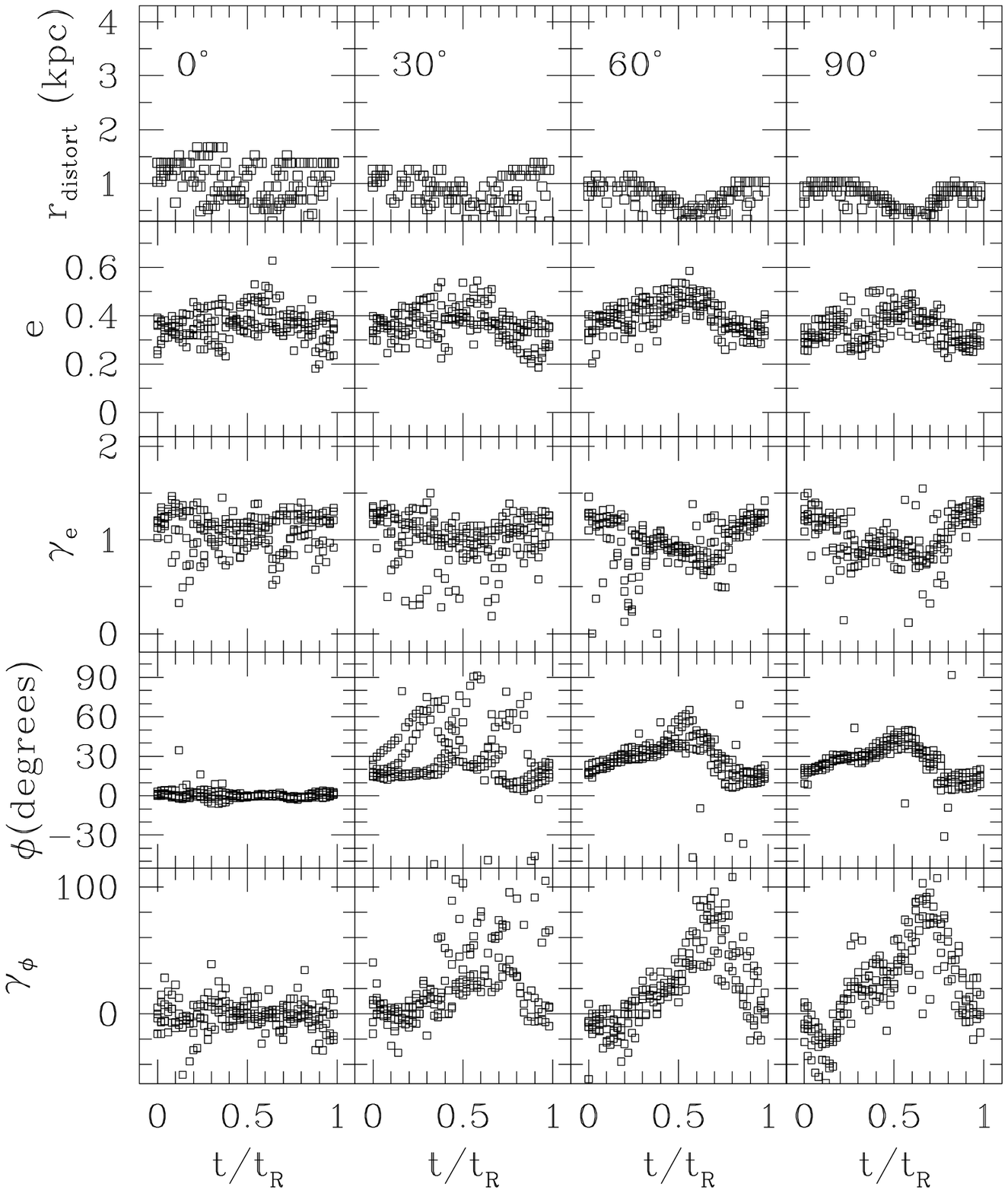}
\caption{Repeat of Figure \ref{ellobs} for Model 2 viewed at
0, 30, 60 and 90 degrees to the orbital plane.
\label{einc}}
\end{center}
\end{figure}

Figures \ref{pinc} and \ref{einc} repeat
Figures \ref{profobs} and \ref{ellobs}, respectively, for Model 2 viewed
at 0, 30, 60 and 90 degrees to its orbital plane
(results were similar for all models).
Since we are interested in satellites of external galaxies,
a single line-of-sight that did not coincide with the view from the
center of the parent galaxy (as might have been more appropriate for 
comparison to Milky Way satellites) was chosen for each viewing angle and 
held fixed for all orbital phases.
The only observable quantities that depend significantly 
on viewing angle are the ellipse orientation parameters $\phi$ and
$\gamma_\phi$ --- all
ellipses align when viewed in the plane of the orbit. 
In a perfectly spherical parent galaxy, debris particles 
will stream within the satellite's
orbital plane, so when viewed within that plane 
there can
be no isophote twisting, but simply elongation along the orbit.
This will be true close
to the satellite even in the case when the parent galaxy is far
from spherical since precession of the orbit will only be apparent 
in debris separated by large orbital phase from the parent.

When we originally performed our analysis
we also found that
$\Sigma_{\rm break}$ and $\gamma_{\rm xt}$ were systematically 
higher when the satellite was viewed from within its orbit.
This effect is most pronounced in Model 1, 
but is somewhat artificial. 
We repeated the analysis, 
specifically excluding all particles at distances more than 15\,kpc
from the satellite along the line of sight, and the differences between
the in-plane and out-of-plane views were smaller.
We concluded that our original analysis was contaminated by debris 
that had stretched entirely around the parent galaxy to intersect the
line of sight.
Clearly, in the case of Galactic satellites this contamination would
only be a problem for satellites within the solar circle.
However, it is possible that it could significantly affect
in-plane views of satellites around other galaxies.

\section{Discussion: Derived Quantities}

\subsection{Parameters of Orbit, Orbital Phase and Viewpoint}

We begin the discussion with a summary of our findings in the previous section
on what surface density maps can tell us about the orbit of the satellite.

\subsubsection{Viewpoint}

Figures \ref{pinc} and \ref{einc} demonstrate that
most observed quantities are not strongly affected by viewing angles
greater than 30 degrees from the orbital plane
of the satellite. 
This result is encouraging because it limits the number of parameters
that typically need to be considered when interpreting
an image.
However, the converse of this statement is 
that it is hard to recover the viewpoint
from imaging data alone.
The one circumstance that we can clearly interpret
in terms of viewpoint is if there is no isophote twisting,
then we are viewing a satellite from
within its own orbital plane. 
The aligned isopleths generated from star count data for
each of the dwarf spheroidal satellites of our own Milky Way 
are a prime example of this case \citep{ih95}.

\subsubsection{Orbital Type and Orbital Phase}

Our results in preceding sections emphasize that the nature of breaks is both
orbit and phase dependent.
However, Figures \ref{profobs} and \ref{ellobs}
also suggest a few rules of thumb in interpreting surface brightness
distributions.

We found that $e, \gamma_e$ and $\phi$ are all much lower at the location of
a break in the brightness profile caused by heating than one caused by
stripping --- or, equivalently, $\rb/\re\sim 1$ for a break caused by
heating.
Such breaks are usually only seen along eccentric orbits and not near
the pericentric phases of these orbits. 
They are accompanied by steep slopes in the surface brightness profiles
of the extra-break population.
Hence, the combination of
a steep slope ($\gamma_{\rm xt} < -3$) and $\rb/\re\sim 1$
is indicative of a non-circular orbit, with the most likely inferred orbital
eccentricity increasing with the steepness of the slope $\gamma_{\rm xt}$.

In contrast, $\rb/\re >1$ and $\gamma_{\rm xt} > -3$
both indicate that we are looking at a break 
where material is being actively stripped. This
occurs at any phase along
a circular orbit, or near pericenter along a more eccentric orbit.

Note that although it is tempting to use $\Delta \gamma$ as a diagnostic 
in the same way as $\gamma_{\rm xt}$ it would be more difficult to
interpret.
We expect $\Delta \gamma$ to depend on the intrinsic profile shape
--- as reflected in the systematic offset between points in the last two
panels of the bottom row of Figure \ref{profobs}, 
In contrast,
$\gamma_{\rm xt}$ depends predominantly on the fractional variance in 
the mass loss rate and
rate of debris dispersal along an orbit, which we expect to be only
loosely dependent on the intrinsic profile shape 
--- as reflected by the much closer agreement between the last two panels 
of the row above.

We conclude that we can place loose constraints on orbital type and
phase using surface brightness data alone.

\subsubsection{Direction of Motion}

Figure \ref{ellobs} shows that though the ellipse orientation and twists
tend to point towards the direction of motion
along near-circular orbits --- i.e., $\phi$ and, to a lesser extent,
$\gamma_\phi$ tend to be positive for Models 1--3 ---
they can be in the opposite sense along more eccentric orbits.  
Hence determination of the direction of motion from surface
brightness data requires some prior constraint on the orbit type
or phase.

\subsection{Relationship of $\re$ and $\rb$ to the tidal
radius}

Since one physical interpretation of 
$\rb$ is the point at which 
the observations become dominated by material leaving the satellite,
it is tempting to think of it as analogous to the {\it tidal
radius\/} $\rt$ of the satellite. 
An estimate of $\rt$ for a satellite provides a way of measuring its
mass, given that of the parent galaxy (or
vice versa) and hence constraining the satellite's dark matter content
\citep{fl83,m96,burkert97}. 
Unfortunately,
Figure \ref{profobs} already suggests that $\rb$ will be
a poor indicator of $\rt$ --- the phase of the orbit
where the satellite is most severely tidally limited and
$\rt$ is smallest is at pericenter, yet 
there is no signature of a corresponding minimum in $\rb$. 
A possible explanation is that $\rb$ reflects the tidal limit
set at pericenter, as was assumed by \citet{k62} in his
derivation of tidal radii and as seen by
\citet{ol92} in their simulations of the disruption of globular clusters.

\begin{figure}
\begin{center}
\epsscale{1.0}
\plotone{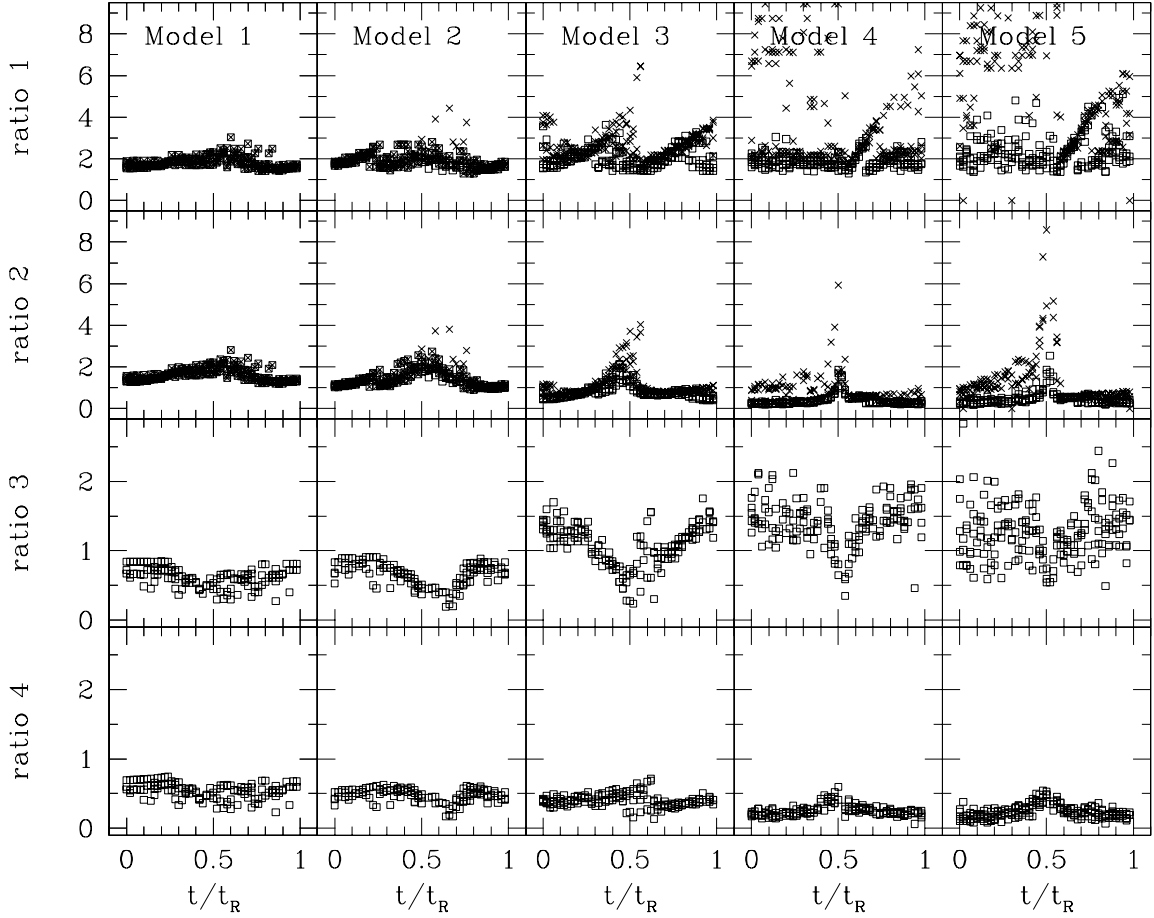}
\caption{Ratio of $\rb$ to $r_{\rm tide, peri}$ (``ratio 1'') and 
$r_{\rm tide, inst}$ (``ratio 2'')
as a function of orbital phase (upper panels)
for $\rb$ defined using the limits
$\Delta \gamma_{\rm lim}=0.2$ (open symbols) and $\Delta \gamma_{\rm lim}=1.0$ 
(crosses). Ratio of $\re$ to $r_{\rm tide, peri}$ (``ratio 3'')
and $r_{\rm tide, inst}$ (``ratio 4'') 
as a function of orbital phase (lower panels).
\label{rt}}
\end{center}
\end{figure}

\begin{figure}
\begin{center}
\epsscale{1.0}
\plotone{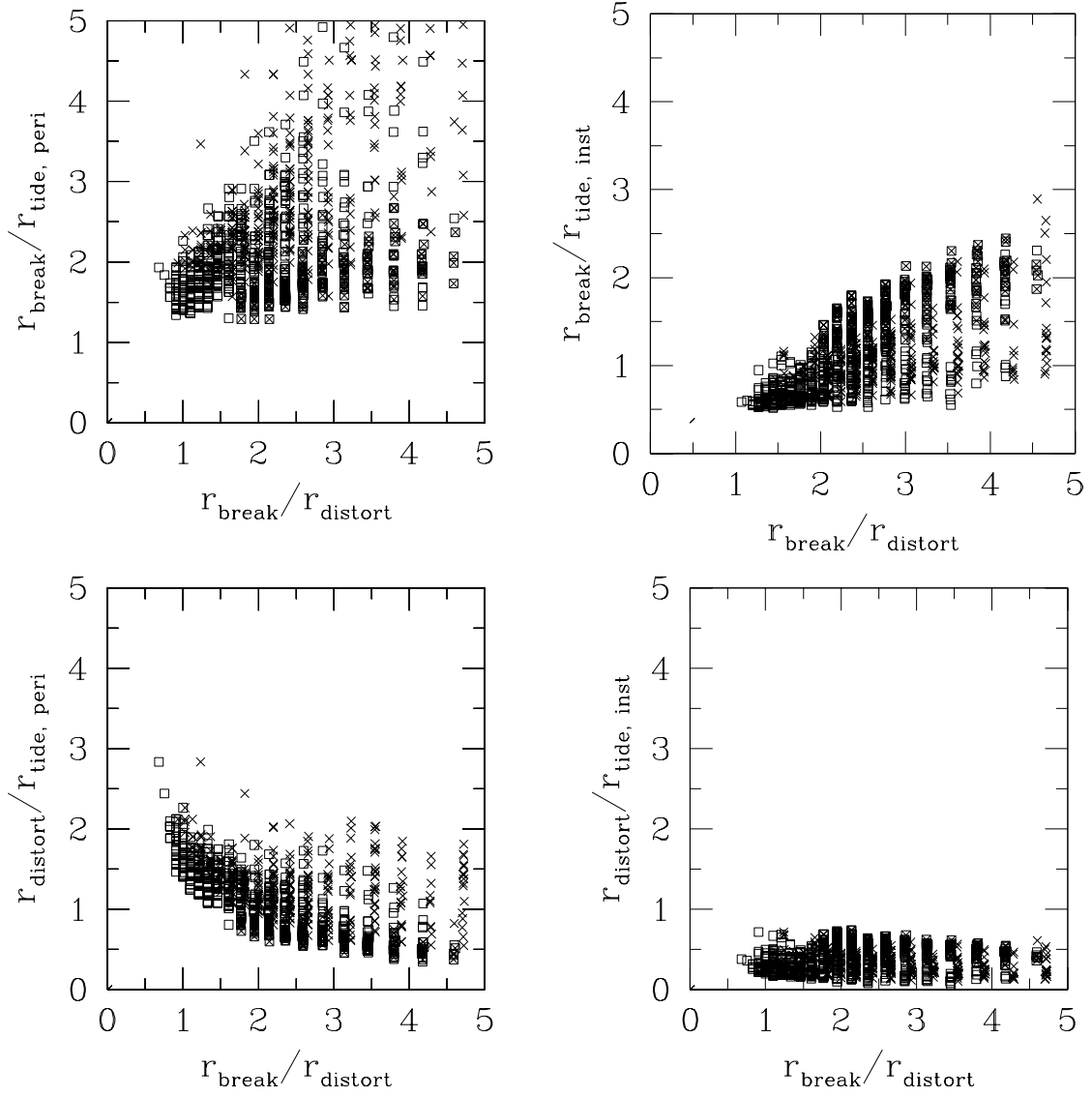}
\caption{Ratio of $\rb$ to $r_{\rm tide, peri}$ and 
$r_{\rm tide, inst}$ (upper panels) as a function of
$\rb/\re$
for $\rb$ defined using the limits
$\Delta \gamma_{\rm lim}=0.2$ (open symbols) and $\Delta \gamma_{\rm lim}=1.0$ 
(crosses). Ratio of $\re$ to $r_{\rm tide, peri}$ and $r_{\rm tide, inst}$ 
as a function of $\rb/\re$ (lower panels).
\label{rbre}}
\end{center}
\end{figure}

\begin{figure}
\begin{center}
\epsscale{1.0}
\plotone{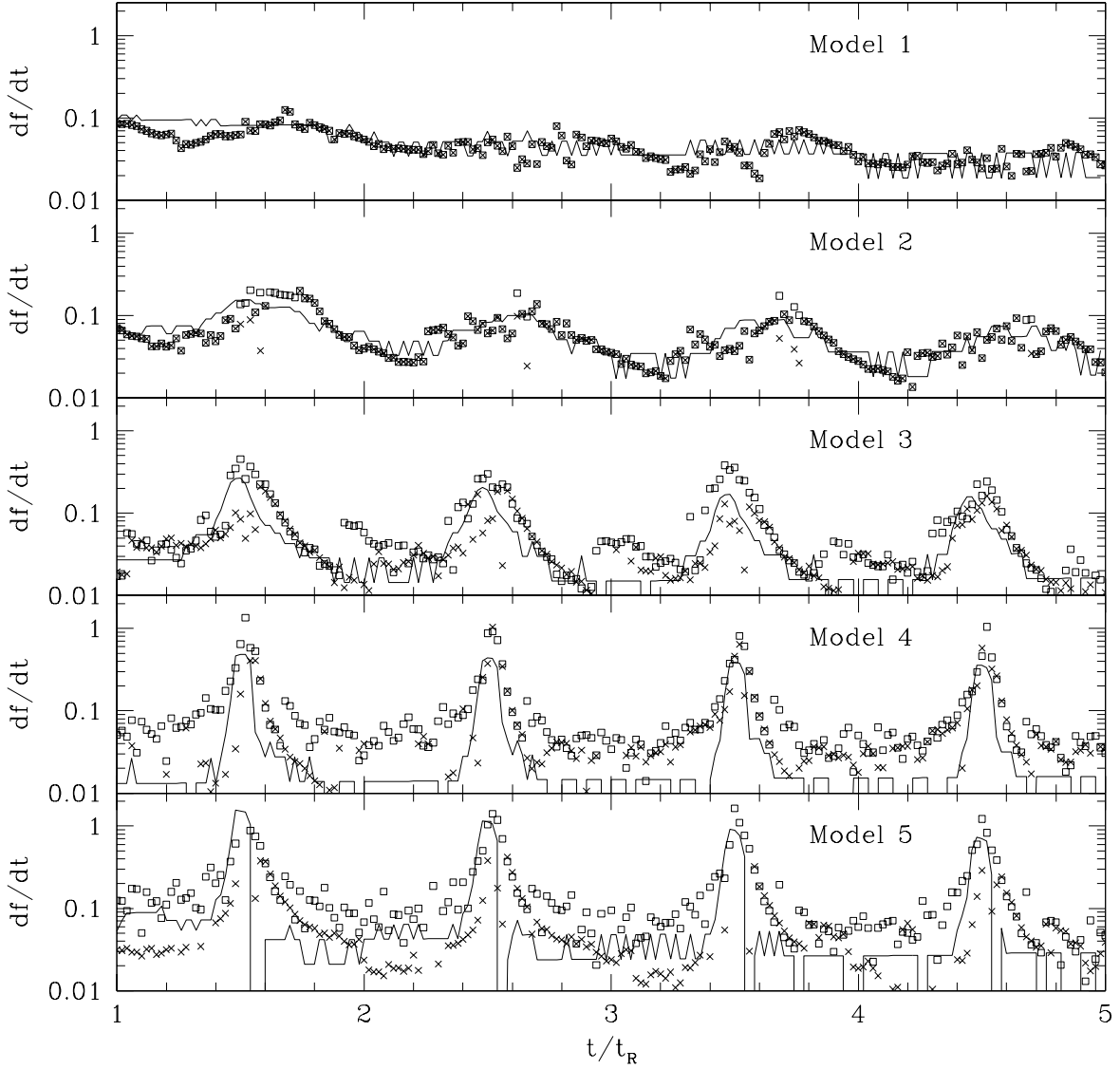}
\caption{The solid lines represent 
the mass loss rate in the simulation and the
boxes
representing the mass loss rate estimated from 
equation (\ref{sigxt2}) for $\rb$ defined using the limits
$\Delta \gamma>0.2$ (open symbols) and $\Delta \gamma > 1.0$ (crosses).
\label{mloss}}
\end{center}
\end{figure}

As a check of this behavior, the top row of panels of 
Figure \ref{rt} shows the ratio of $\rb$
to the tidal radius set at pericenter,
$\rtp$ (labeled as ``ratio 1''), calculated from King's formula,
\begin{equation}
	r_{\rm tide, peri}= R_{\rm peri} \left({
m_{\rm sat} \over M_{\rm gal, peri} (3+e_{\rm orb})}\right)^{1/3}
\label{rtide}
\end{equation}
using the known values of the mass of the satellite
$m_{\rm sat}$, pericenter of the satellite's orbit $R_{\rm peri}$, 
mass of the parent galaxy enclosed within this distance $M_{\rm gal, peri}$ 
and eccentricity of the satellite's orbit $e_{\rm orb}$
for each simulation.
(Note: for $m_{\rm sat}$ we tried using both the mass enclosed in projection
within the $\rb$ isophote and the mass bound to the satellite
from the simulations. We found the latter to be typically 10\% larger
than the former, which would change $\rtp$ by only 3\%.)
The open squares represent $\rb$ (defined using $\Delta\gamma_{\rm lim}=0.2$
as before), and the crosses are for the analysis repeated, this time
requiring $\Delta \gamma_{\rm lim}=1$ (and with a grid extending to
40\,kpc$\times$40\,kpc for Models 3--5).
Recall that along the more eccentric orbits breaks found using 
$\Delta\gamma_{\rm lim}=0.2$ tended to reflect where the bound
portion of the satellite had merely been distorted by the tidal field, while 
those found using the larger 
of these limits corresponded to the the point where stripping
was actively occurring (see \S 3.2.1).
In all the models, $\rb > \rtp$ and
the open squares hover around $\rb/\rtp \sim 2$ throughout
each orbit, suggesting that this break is indeed in response to 
a fundamental limit set at pericenter. 
The open squares and crosses are coincident for Models 1 and 2, but the
second set occur at distinctly larger radii in Models 3--5 --- a reflection
of the separation of the radius where stripping is occurring from where
the satellite has actually been distorted by the tidal field, described above.

The second row of panels in Figure \ref{rt}
show the ratio of $\rb$ to the instantaneous
tidal radius $\rti$ (``ratio 2'') for a satellite on a circular orbit at the
current distance of the satellite from the parent galaxy $R$ (i.e.
replacing $R_{\rm peri}$ with $R$, $M_{\rm Gal, peri}$ with
$M_{\rm Gal}$ and setting $e_{\rm orb}=0$ in
equation [\ref{rtide}]).
The third and fourth row of panels repeat the first and second rows for the
ratio of $\re$ to $r_{\rm tide, peri}$ (``ratio 3'') and
$r_{\rm tide, inst}$ (``ratio 4'') .  From
the second row we can see that the inner and outer breaks straddle 
$\rti$ for much of the orbit for Models 3--5, while the fourth row shows
$\re < \rti$ for each model and
throughout each orbit.
These characteristics
emphasize that it can be possible to see elongations, twists and
break radii well within the
point where one might expect tidal effects to be
important (e.g. by calculating $\rti$ using dynamical estimates for
$M_{\rm gal}$ and $m_{\rm sat}$).

We have already concluded that $\rb/\re$ can place weak constraints on
the orbit type and phase, and indicate whether a break is due to
heating or stripping processes.
Figure \ref{rbre} shows $\rb/\rt$ and $\re/\rt$, calculated using both
pericentric and instantaneous values of $\rt$, as a function
of $\rb/\re$.
It is intended to illustrate the extent to which the observables $\rb$ and
$\re$ might be used to constrain $\rti$ and $\rtp$.
We find:
(i) if $\rb/\re\sim 1$ then $\rtp < \rb < 2 \rtp$ 
and we expect to be at a phase where $\rti > \rtp$;
and (ii) if $\rb/\re>2$ then $0.5 \rtp
< \re <1.5 \rtp$ and $\rti < \rb < 2\rti$.
Unfortunately, 
such arguments can at best lead to an
estimate of the tidal radius that is good to within a factor of two. 
In these circumstances 
it is dangerous to use our estimated $r_{\rm tide}$ in
equation (\ref{rtide}) to estimate 
the mass of either the parent or satellite galaxy 
since this order unity error in $r_{\rm tide}$ would lead
to an order of magnitude error in the mass estimate.
Indeed, it may be possible to reconcile the small cutoff radii observed
when deriving surface density profiles of the Milky Way's dwarf spheroidal
satellites from star-counts \citep{ih95} with the large dark matter content
implied by their velocity dispersions by assuming these satellites are
near the apocenters of their highly-eccentric orbits.
Such a conjecture needs to be tested more thoroughly with models
specifically designed to mimic the observed properties of these galaxies.

\subsection{Mass Loss Rates}

In their interpretation of debris populations, \citet{jsh99} 
derived an expression for the surface density of the 
extra-break population by
assuming that once material is unbound from a satellite
it is not influenced significantly by the satellite's gravitational field and
by arguing that debris material
will spread away from a satellite at the characteristic rate
$\pi \rb/ T_{\rm orb}$.
They found
\begin{equation}
	\Sigma_{\rm xt}(r)={dm \over dt} {T_{\rm orb} \over \pi}
{1 \over 2 \pi \rb r}
\label{sigxt1}
\end{equation}
where $T_{\rm orb}$ is the azimuthal time period of the satellite's
orbit (time taken to cover 2$\pi$ in angle along the orbit)
and $dm/dt$ is the mass loss rate from the satellite.  By
overlaying this expression on the surface density profiles in the
simulations they demonstrated that it compared well for satellites on
near-circular orbits, and concluded that outlying populations with
slopes of $\gamma=-1$ are likely to be explained as tidal debris.  By
contrast, note that $\gamma$ was typically less than $-1$ in all our
simulations (Figures \ref{profobs} and \ref{pinc}), and in those of
\citet{clm99}.

The apparent contradiction between our simulations and
equation (\ref{sigxt1})
can be resolved by noting that: 
(i) close to the
satellite we do expect debris particles
to still be influenced
to some extent
by tidal forces (in contrast to
the assumption in the previous paragraph)
--- in Figure \ref{snaplil} the particles are
seen to move perpendicular to the satellite's orbit
(i.e., in the direction of tidal forces)
before being stretched along it; 
(ii) both the mass loss rate and the rate of debris
dispersal are phase-dependent along eccentric orbits; 
(iii) breaks can occur well within the still-bound population of
the satellite.

An alternative expression for the
surface density that captures some of the above behavior is
\begin{equation}
	\Sigma_{\rm xt}(r) \propto {dm \over dt} {T_{\rm circ} \over r^2}
\label{sigxt2}
\end{equation}
where the drift rate is now assumed
to increase with $r$ and
the variable rate of debris dispersal along the orbit 
has been represented by replacing $T_{\rm orb}$ with
the time period for a circular orbit at
the satellite's instantaneous position, $T_{\rm circ}$.
In fact, throughout Model 1 (the near-circular case) the slope
hovered in the range $-3 <\gamma_{\rm xt} < -1$, suggesting that
the true behavior lies somewhere between the simplified representations
given in equations (\ref{sigxt1}) and (\ref{sigxt2})
as debris stars are observed in transition between being bound and
unbound.
This effect was not noted by \citet{jsh99} because they did not
explicitly fit slopes to their extra-tidal populations but rather overlaid
models on the surface density profiles.  Also, they did not consider
eccentric orbits where the variance in slopes is even more 
pronounced and only examined their simulations at a limited number
of orbital phases.
(Note that neither approach takes into account the variable mass loss rate 
along the orbit which would also affect $\gamma_{\rm xt}$ and could explain 
the more extreme $\gamma_{\rm xt}<-2$ cases not consistent with
either equation.)

\citet{jsh99} re-arranged equation (\ref{sigxt1}) to estimate the mass loss 
rate from a satellite using observed quantities, and found that,
when applied to simulations of satellites on near-circular orbits, 
this approach could be used to reproduce the mass loss rate
in the simulation to within a factor of two.
In Figure \ref{mloss} we re-examine the accuracy of this approach 
at all phases of the more eccentric orbits studied in this paper,
using equation (\ref{sigxt2}) evaluated at $\rb$.
We first find the ``best'' constant of proportionality for this
equation to be $5/(4 \pi^2)$ by requiring the estimate to closely match
the known mass loss rate in the simulation throughout Model 1 (where
the $\pi^2$ is used instead of a numerical value to mimic equation 
[\ref{sigxt1}]).
We then use this value for our estimates along the more eccentric
orbits.
The solid line represents the mass loss rate in the simulation (averaged
over one tenth of the orbit) and the
symbols are the ``observed'' mass loss rate estimated at $\rb$, defined as
usual with $\Delta\gamma_{\rm lim}=0.2$ (open squares), and at
$\rb$ defined with $\Delta \gamma_{\rm lim}=1.0$ (crosses).
The mass loss estimate derived from observable quantities
reproduces the approximate amplitude as a function
of phase along Models 1--3.
It is less successful along the most eccentric orbit, in particular where
we know the break to be within the bound satellite population
and do not expect the debris to be dispersing freely.

Note that this idea can be extended once longer portions of
a satellite's debris stream are available to reconstruct 
a detailed mass loss history from breaks between
populations lost on different orbits or different disk passages 
(see Combes, Leon \& Meylan, 1999 and Leon, Meylan \& Combes, 2000; also
Johnston et al., 1999, for a discussion
of this idea in the context of the Sagittarius dwarf spheroidal galaxy).

\section{Summary}

In this paper we have attempted to understand how the 
morphology of faint populations around
satellite galaxies might be interpreted in terms of the orbit
and orbital phase of the satellite, and the tidal
influence of the parent galaxy.
We have found that
breaks in the surface density profile can be signatures
of heating within the bound satellite population,
stripping of debris from the satellite, or the transition between
distinct debris populations.
Multiple breaks can be found within a single satellite profile and 
attributed to one or more of these effects.
Breaks can be above or below the inner profile of the
satellite and the surface density of extra-break material can
have widely varying amplitude and slopes (we found $-5<\gamma_{\rm xt}<-1$).
Morphological distortions to the satellite in the form of
increasing ellipticity and twisting of isophotes can occur
well within any identified break radii.
The identification of break radii depends on the
choice of $\Delta \gamma_{\rm lim}$,
inner satellite profile and resolution of observations.

Using surface brightness distributions alone we expect to be able
place loose constraints on either the preicentric or instantaneous 
tidal radius, mass loss rate, 
orbital type and phase of the satellite and nature of breaks in the
surface brightness profile.
For example, we have found that breaks well within the bound satellite 
population tend to only occur along the more eccentric orbits
at phases when the satellite is not near pericenter.
Such breaks are usually coincident with the onset of other
morphological disturbances such as increasing 
isophote ellipticity or twisting (or $\rb \sim \re$).
By contrast, along circular orbits and near pericenter of eccentric orbits
breaks tend only to be identified at radii much greater than where
disturbances in the morphology first become apparent (or $\rb > \re$ ---
note that this can also occur at apocenter along eccentric orbits if
the inner break is too subtle to be detected) and usually indicate the
region where material is being actively stripped from the satellite.
Hence a satellite with $\rb \sim \re$ is likely to be on an 
eccentric orbit and have an instantaneous tidal radius much larger than $\rb$.
One with $\rb > \re$ could be at any phase of a circular or eccentric
orbit, with the break corresponding to the transition between bound and
unbound material.
Our results are largely independent
of viewpoint, as long as we are not viewing the satellite
directly within its orbital plane.

We conclude that though
surface brightness distributions contain a wealth of information,
degeneracies in orbital type and orbital phase dependencies
limit the extent to which they can be interpreted physically.
In a further paper we look at what more might be gained by
using photometric data in conjunction
with spectroscopic data.

\acknowledgements 
We thank the referee for her/his many detailed and insightful
comments.  KVJ acknowledges support in part as a member of the
Institute for Advanced Study, and from NASA LTSA grant NAG5-9064.  PIC
thanks the ARCS foundation and the NSF for support as an ARCS
Foundation scholar and an NSF graduate student research fellow.

\begin{thebibliography}{}

\bibitem[Binney \& Tremaine(1987)]{bt87}
Binney, J. \& Tremaine, S. 1987, {\it Galactic Dynamics} 
(Princeton: Princeton University
Press)

\bibitem[Burkert(1997)]{burkert97} 
Burkert, A. 1997, \apjl, 474,
L99 

\bibitem[Choi, Guhathakurta \& Johnston(2002)]{paper2}
Choi, P. I., Guhathakurta, P. \& Johnston, K. V. 2002, \aj, submitted

\bibitem[Combes, Leon \& Meylan(1999)]{clm99}
Combes, F., Leon, S. \& Meylan, G. 1999, \aap, 352, 149

\bibitem[Faber \& Lin(1983)]{fl83}
Faber, S. M. \& Lin, D. N. C. 1983, \apjl, 266, L17

\bibitem[Ghigna et al(1998)]{ghigna98}
Ghigna, S., Moore, B., Governato, F., Lake, G., Quinn, T., \& Stadel, J. 1998,
\mnras, 300, 146

\bibitem[Gould et al.(1992)]{ggrf}
Gould, A., Guhathakurta, P. Richstone, D. \& Flynn, C. 1992, \apj,
388, 354

\bibitem[Grillmair(1998)]{g98}  
Grillmair, C. J. 1998, in {\it Galactic Halos}, ed. D. Zaritsky
(ASP Conf. Ser., Vol. 136), p. 45

\bibitem[Grillmair et al.(1996)]{g+96}  
Grillmair, C. J.,  Lauer, T. R., Worthey, G.,
 Faber, S. M., Freedman, W. L.,
 Madore, B. F., Ajhar, E. A., Baum, W. A.,
 Holtzman, J. A., Lynds, C. R.,
 O'Neil, E. J., Jr. \& Stetson, P. B. 1996, \aj, 112, 1975

\bibitem[Grillmair et al.(1995)]{g+95}  
Grillmair, C. J., Freeman, K. C., Irwin,
M. \& Quinn, P. J. 1995, \aj, 109, 2553

\bibitem[Guhathakurta, Choi \& Raychaudhury(2002)]{gcr01}
Guhathakurta, P., Choi, P. I. \& Raychaudhury, S. 2002, \aj, in prep

\bibitem[Helmi \& White(1999)]{hw99}
Helmi, A. \& White, S. D. M. 1999, \mnras, 307, 495

\bibitem[Hernquist(1990)]{h90}
Hernquist, L. 1990, \apj, 356, 359

\bibitem[Hernquist \& Ostriker(1992)]{ho92}  Hernquist, L. \& Ostriker,
J. P. 1992, \apj, 386, 375

\bibitem[Hodge(1973)]{hodge73}
Hodge, P. W. 1973

\bibitem[Irwin \& Hatzidimitriou(1995)]{ih95}
Irwin, M. J. \& Hatzidimitriou, D. 1995, \mnras, 277, 1354

\bibitem[Johnston(1998)]{j98}
Johnston, K. V. 1998, \apj, 495, 297

\bibitem[Johnston et al.(1999)]{j+99}
Johnston, K. V., Majewski, S. R.,
Siegel, M. H., Kunkel, W. E.
\& Reid, I. N. 1999, \aj, 118, 1719

\bibitem[Johnston, Hernquist \& Bolte(1996)]{jhb96} 
Johnston, K. V., Hernquist, L. \& Bolte, M. 1996, \apj, 465, 278 

\bibitem[Johnston, Sigurdsson \& Hernquist(1999)]{jsh99}
Johnston, K. V., Sigurdsson, S \& 
Hernquist, L. 1999, \mnras, 302, 771 

\bibitem[Kent(1987)]{kent87}
Kent, S. M. 1987, \aj, 94, 306

\bibitem[King(1962)]{k62} 
King, I. R. 1962, \aj, 67, 471

\bibitem[Klessen \& Kroupa(1998)]{kk98} 
Klessen, R.~S.~\& Kroupa, P.\ 1998, \apj, 498, 143

\bibitem[Kuhn(1993)]{k93} 
Kuhn, J.~R.\ 1993, \apjl, 409, L13

\bibitem[Kuhn \& Miller(1989)]{km89} 
Kuhn, J.~R.~\& Miller, R.~H.\ 1989, \apjl, 341, L41

\bibitem[Kuhn, Smith \& Hawley(1996)]{ksh96}
Kuhn, J. R., Smith, H. A. \& Hawley, S. L. 1996, \apjl, 469, L93
 
\bibitem[Leon, Meylan \& Combes(2000)]{lmc00}
Leon, S., Meylan, G. \& Combes, F. 2000, \aap, 359, 907

\bibitem[Majewski et al.(2000)]{m+00}
Majewski, S. R.,  
Ostheimer, J. C.,
Patterson, R. J.,
Kunkel, W. E.,
Johnston, K. V. \& Geisler, D. 2000, \aj, 119, 760 

\bibitem[Mateo, Olszewski \& Morrison(1998)]{mom98}
Mateo, M., Olszewski, E. W. \& Morrison, H. L. 1998, \apjl, 508, L55 

\bibitem[Miyamoto \& Nagai(1975)]{mn75}
Miyamoto, M. \& Nagai, R. 1975, PASJ, 27, 533

\bibitem[Moore(1996)]{m96}
Moore, B. 1996, \apjl, 461, L13

\bibitem[Odenkirchen et al.(2000)]{sdss}
Odenkirchen, M., et al. 2000, astro-ph/0012311

\bibitem[Oh \& Lin(1992)]{ol92}
Oh, K. S. \& Lin, D. N. C. 1992, \apj, 386, 519

\bibitem[Quinn(1984)]{q84}
Quinn, P. J. 1984, \apj, 279, 596

\bibitem[Toomre \& Toomre(1972)]{tt72}
Toomre, A. \& Toomre, J. 1972, \apj, 178, 623


\bibitem[Tormen, Bouchet, \& White(1997)]{tormen97} 
Tormen, G., Bouchet, F.~R., \& White, S.~D.~M.\ 1997, \mnras, 286, 865

\bibitem[Tremaine(1993)]{t93}
Tremaine, S. 1993, in {\it Back to the Galaxy}, eds. S. S. Holt \& F. Verter
(New York: AIP), p. 599

\bibitem[Weinberg(1994a)]{w94a} 
Weinberg, M.~D.\ 1994, \aj, 108, 1398

\bibitem[Weinberg(1994b)]{w94b} 
Weinberg, M.~D.\ 1994, \aj, 108, 1414

\bibitem[Weinberg(1994c)]{w94c} 
Weinberg, M.~D.\ 1994, \aj, 108, 1403

\end{thebibliography}
\end{document}